\documentclass[useAMS,usenatbib]{mn2e}

\usepackage{epsfig,aas_macros}

\title[The structure of the Galactic bar]{
The structure of the Galactic bar\thanks{Based on
observations with CIRSI, the Cambridge InfraRed Survey Instrument,
obtained at the duPont 2.5m telescope of Las Campanas
Observatory}}
\author[C. Babusiaux and G. Gilmore]
 {C.~Babusiaux$^{1,2}$ and G.~Gilmore$^1$\\
$^1$ Institute of Astronomy, University of Cambridge, Cambridge, CB30HA, UK\\
$^2$ Institut d'Astronomie et d'Astrophysique, Universit\'e Libre de Bruxelles, B-1050 Bruxelles}
\begin{document}

\date{Accepted xxx. Received xxx}

\pagerange{\pageref{firstpage}--\pageref{lastpage}} \pubyear{2004}

\maketitle

\label{firstpage}

\begin{abstract}
We present a deep near-infrared wide-angle photometric
analysis of the structure of the inner Galactic bar and central disk.
The presence of a triaxial structure at the centre of the Galaxy is
confirmed, consistent with a bar inclined at 22$\pm$5.5\degr\ from
the Sun-Galactic centre line, extending to about 2.5~kpc from the
Galactic centre and with a rather small axis ratio. A feature at
$\ell$=$-9.8$\degr\ not aligned with this triaxiality suggests the existence
of a second structure in the inner Galaxy, a double triaxiality or an inner
ring. We argue that this is likely to be the signature of the end of
the Galactic bar, at about 2.5-3~kpc, which is circumscribed by an inner
pseudo-ring.  No thick dust lane preceding the bar is detected, and a
hole in the disc's dust distribution inside the bar radius is
inferred.
\end{abstract}


\section{Introduction}
The central region of the Galaxy is our closest opportunity to
quantify the structure and formation of a dense and massive complex
stellar-dynamical system. A detailed star-by-star analysis is possible
in the Milky Way bulge, in contrast to distant galaxies where only the
integrated light is observed.  But our point of view from within the
Galactic disc has its drawbacks. The high interstellar extinction, the
crowding and the confusion between foreground disc stars and bulge
sources, make studies of the inner Galactic regions difficult.

There is substantial evidence for the presence of a triaxial structure
in the inner Galaxy. The presence of a bar was first suggested from
the observations of non-circular motions in the gas kinematics by
\cite{DEV64}. The first direct evidence came from detection of an
asymmetry in the infrared luminosity distribution (\citealt{BLI91}),
confirmed by the COBE map (\citealt{DWE95}). Star counts confirmed the
presence of an asymmetry in their luminosity distribution, from IRAS
sources (\citealt{NAK91}, \citealt{WEI92}), near-infrared star surveys
(\citealt{HAM94}, \citealt{LOP97}, \citealt{UNA98}), and optical
(OGLE) surveys of red clump stars (\citealt{STA94}).  The new
near-infrared surveys DENIS and 2MASS, with arcsecond spatial
resolution and sufficient sensitivity to detect typical red giant
bulge sources, complemented by the mid-infrared ISO survey ISOGAL, are
beginning to provide new constraints (\citealt{LOP01},
\citealt{COL02}, \citealt{VAN03}). The high optical depth measured by the
microlensing surveys also points towards a non-axisymmetric model, but
the exact constraints remain controversial (e.g. \citealt{BIN00}).  

Only a few dynamical constraints are available so far, essentially
from stellar kinematics from SiO masers (\citealt{DEG00}), OH/IR stars
(\citealt{SEV99}), and from observations through a few low-extinction
windows (\citealt{HAF00}).  Early dynamical models of gas HI, CO and
CS emissions (\citealt{BIN91}) and infrared luminosity distribution
models (\citealt{DWE95}, \citealt{FRE98}) were derived independently,
while more recently new models are being developed to incorporate both
gas and luminosity constraints, using in particular N-body simulations
(\citealt{FUX99}).

All those studies agree that the inner Galaxy at low latitudes is
asymmetric, with the bulge/bar major axis at positive galactic longitudes. 
The exact
orientation, length and axis ratio of the inner asymmetric structure
are still poorly constrained (\citealt{GER01}).  The relative
importances of the bulge and the inner disk in the central regions,
the stellar population content, and the formation and evolutionary
history remain unknown (\citealt{WYS97}). Discussion continues on the
need for a distinction between a central bulge and a long thin bar
(\citealt{KUI96}). Indeed, along the line of sight towards the central
regions of the Galaxy one crosses the galactic disc, spiral arms, the
molecular ring, the bar and/or bulge, and maybe other complex
structures such as a hole in the disc distribution, or a possible
stellar ring or double bar, as observed in external galaxies.  To
improve our understanding of this crucial region new high sensitivity
and high spatial resolution multi-wavelength studies are needed.

In this paper we present new deep near-infrared photometry of the
inner Galaxy close to the Galactic plane, taken with
the Cambridge InfraRed Survey Instrument (CIRSI). Those observations
provide large samples of bulge/bar red clump stars, allowing
extinction mapping and reliable 3-D spatial analyses in fields at both
positive and negative galactic longitudes in the Galactic plane.

\section{CIRSI photometry of the inner Galaxy}

The requirements for an observational study of inner Galactic
structure are well-established, following earlier studies (e.g.
\citealt{UNA98}, \citealt{IBA95a}, \citealt{IBA95b}). 
Near infrared observations
are optimal, to identify the dominant intrinsically-bright red giant
stellar population, while minimising the effects of interstellar
extinction (longer wavelength mid-infrared studies, such as those for
ISOGAL, are optimised for study of the latest M giants).  Observations
of the inner Galaxy preferably in the Galactic plane, and certainly at
low Galactic latitudes, are desirable to minimise model-dependent
corrections for the as-yet poorly known scale heights of the
inner-galactic stellar populations.  To study the expected asymmetry
caused by the Galactic bar, observations at both positive and negative
longitudes are essential, with, in so far as is possible given the
extinction distribution, symmetric longitude fields
preferred. \cite{UNA98} have shown that structural studies at absolute
galactic longitudes greater than 4 degrees are needed to constrain bar
models to a useful extent.

Given these requirements, we selected for further detailed study
fields observed by DENIS and ISOGAL at galactic latitudes
$-0.3\degr<b<0.3\degr$ and galactic longitudes $\ell=\pm5-6\degr$,
where the effect of any bar-like feature should be strong, and at
$\ell$=$\pm9-10\degr$ where the bar may or may not be present.  The
field C32 ($\ell=0\degr$,$b=1\degr$), which was used as an ISOGAL
calibration field because of its low and uniform extinction
(\citealt{OMO99}), was also selected for calibration purposes here,
and in addition to provide the minor-axis reference to set the
zero-point of the expected asymmetry.

Near-infrared observations in the J(1.25$\mu$m), H(1.65$\mu$m) and
K$_s$(2.15$\mu$m) bands were obtained with the Cambridge Infrared
Survey Instrument (\citealt{BEC97}, \citealt{MAC00}) on the du Pont
2.5m telescope at Las Campanas Observatory.  CIRSI is a mosaic imager consisting
of four Rockwell 1K x 1K detectors. The pixel scale is 0.2 arcsec
pixel$^{-1}$.  The gaps between the detectors being comparable to the
detector size, four dither sets are used to create a fully-observed mosaic image,
leading to a basic field of view of about 13 x 13 arcmin$^2$.

The observations carried out for this survey are summarised in table
\ref{ObsLog}. 

\begin{table*}
\caption{Log of the CIRSI observations. 
Positions of the centre of the mosaic are given both in galactic and
equatorial J2000 coordinates.
The exposure mode is given by the number
of dither frames x the number of sub-integrations(loops) with their individual exposure time. ($^1$) after the exposure time indicates that chip~3 has been discarded, see footnote$^1$.
The seeing conditions are indicated by the PSF FWHM with its
variation range during the mosaic observation. The last column
indicates the effective useful magnitude limit.}
\label{ObsLog}
\begin{tabular}{ll|lllll}
\hline
Field ($\ell$,$b$) & RA (J2000) & Filter & Date & Exposure & seeing & mag limit \\
 & DE (J2000) & & & & (\arcsec) & (mag) \\
\hline
\hline
5NN ($-$5.73,$-$0.22) & 17:32:08.0 & J & 2000-08-17 &
9 x 5*20s & 0.6-0.64 & 21.0 \\
 & $-$33:54:00.0 & H & 2001-09-06 & 9 x 5*20s ($^1$) & 0.72-0.84 & 18.3 \\
 & & K$_s$ & 2001-04-10 & 9 x 3*20s & 0.68-0.7 & 18.3 \\
\hline
5NP ($-$5.74,+0.19) & 17:30:30.0 & J & 2000-08-17 &
9 x 5*20s & 0.66-0.72 & 20.9 \\
 & $-$33:41:00.0 & H & 2000-08-16 & 9 x 5*20s & 0.66-0.82 & 19.5 \\
 & & K$_s$ & 2001-04-14 & 9 x 2*30s & 0.6-0.68 & 19 \\
\hline
5PN (+5.67,$-$0.28) & 17:59:35.0 & J & 2000-08-17 &
9 x 5*20s & 0.8-0.96 & 20.2 \\
 & $-$24:12:00.0 & H & 2000-08-16 & 9 x 5*20s & 0.6-0.76 & 20.2 \\
 & & K$_s$ & 2001-04-10 & 9 x 3*20s & 0.66-0.9 & 18.2 \\
\hline
5PP (+5.76,+0.23) & 17:57:53.0 & J & 2001-09-04 &
9 x 5*20s & 1.04-1.24 & 19.6 \\
 & $-$23:52:00.0 & H & 2000-08-16 & 9 x 5*20s & 0.76-0.8 & 19.3 \\
 & & K$_s$ & 2001-04-15 & 9 x 2*30s & 0.68-0.84 & 18.4 \\
\hline
\hline
9N ($-$9.80,+0.05) & 17:19:51.6 & J & 2001-09-07 &
9 x 5*20s ($^1$) & 1.04-1.2 & 19.7 \\
 & $-$37:07:21.9 & H & 2001-09-08 & 9 x 5*20s ($^1$) & 0.8-0.84 & 18.9 \\
 & & K$_s$ & 2001-09-03 & 9 x 5*20s & 0.8-0.86 & 18.6 \\
\hline
9P (+9.55,$-$0.09) & 18:07:09.6 & J & 2001-09-04 &
9 x 5*20s & 1.4-1.7 & 19.4 \\
 & $-$20:43:28.5 & H & 2001-09-06 & 9 x 5*20s ($^1$) & 0.86-0.96 & 18 \\
 & & K$_s$ & 2001-09-03 & 9 x 5*20s & 1.06-1.4 & 17.9 \\
\hline
\hline
C32 (+0.00,+1.00) & 17:41:45.0 & J & 2001-09-07 &
9 x 5*20s ($^1$) & 0.94-1.1 & 18.1 \\
 & $-$28:25:00.0 & H & 2001-09-08 (3/4) & 9 x 5*20s ($^1$) & 1.06-1.3 & 17.7 \\
 & & H & 2001-10-02 (1/4) & 9 x 5*20s ($^1$) & 0.86 &  \\
 & & K$_s$ & 2001-09-05 & 9 x 5*20s ($^1$) & 0.9-1.1 & 16.6 \\
\hline
\end{tabular}
\end{table*}

\subsection{Data reduction}

The data reduction was carried out using an extensively updated
version of the InfraRed Data Reduction (IRDR) software package, first
developed by \cite{SAB01}. A summary of the full process is given
here. The updated version of IRDR with its documentation was developed
by one of us (CB) and is available at http://www.ast.cam.ac.uk/\~{}optics/cirsi/software.

First each image is corrected for non-linearity, as the array
detectors used are non-linear in their response to flux.  To calibrate the
non-linearity, domeflats were taken with different exposure times,
with a short exposure observation between each as reference. The
expected flux for an exposure time is computed from the reference
exposure flux level.  The ratio of this flux to the measured flux is
observed to increase quadratically with the flux level. A linear
quadratic regression is computed up to the saturation level (about
40,000 ADU), setting the first coefficient to unity to constrain no
correction for a zero signal level. This relation is then used to
correct the non-linearity on all images.

The detectors generate an internal (thermal noise) signal even when
there is no external signal. This `dark' signal must then be
subtracted\footnote{During the observations made at the end of September and during October 2001, a nitrogen leak from the dewar, with consequent
temperature drifts, affected the stability of the dark count and the
linearity of the chip~3 detector so badly that those data had to be
discarded.}. This is a straightforward process, using exposures of the
same length as the science exposures, but with the shutter closed.

The data are then flatfield corrected. This process uses the 
difference image derived by subtracting domeflats obtained with the
illumination lamp turned on from subsequent domeflats obtained with
the illumination lamps turned off. This difference image is normalized
to the sensitivity of the first detector.  These flatfields are also used to
detect bad pixels and to create weight maps, which are used to create a
signal-to-noise value for each pixel observation for use during the
coaddition of all the individual `dither' exposures into a final
single combined image.

The sky is subtracted in two passes. The basic observational `unit' is
a set of typically five 20-second repeated exposures (`loops') on a
single field. Nine of these sets of exposures are obtained, with small
telescope pointing offsets (`dither') between each set. The five loops are 
combined, using sigma clipping, into a single frame called `dither frame'. 
A first-pass sky image is derived by median-combining the nearest dither 
frames. This process of course leaves the real sources in the frame. 
The nine dither frames are then offset
and a preliminary combination is made. After this first dither
frame coaddition, object masks are produced using SExtractor source
extraction (\citealt{BER96}) and used to calculate a mask which can
exclude all detected sources from the raw data frames. Object-masked
frames are used to make a second pass sky subtraction on each basic
loop image.

The spatial offsets generated by the dithering between the 
dither frames are computed by cross-correlating
object pixels detected by SExtractor. The nine individual dither frames are 
then coadded using a weighted bi-linear interpolation, excluding bad pixels.

Finally, the astrometry is calibrated by correlating the SExtractor's 
object catalogue with the 2MASS catalogue.

In practice, the image PSF is seen to vary significantly during the
typically 15 minutes which are required to create a full set of dithered
observations. Consequently, to maximise the photometric quality of the
data, no attempt was made to create single full mosaic images.

\subsection{Point source extraction and photometry}

The borders of the images, which do not have all the dither
observations and so have highly variable signal-to-noise ratios, 
are removed before beginning the source extraction procedure.

PSF-fitting photometry was carried out with the IRAF DAOPHOT package.
A first list of detections was created by the DAOFIND procedure with a
5$- \sigma$ threshold.  An initial estimate of the photometry is then
provided by the PHOT procedure, adopting as aperture the Full Width at Half Maximum (FWHM) value calculated from the image Point Spread Function (PSF) 
(see Table \ref{ObsLog}).
A list of bright and `almost' isolated stars, preselected by the
PSTSELECT routine, was interactively confirmed.  Those stars are used
by the PSF task to compute a PSF model, using a Gaussian profile and a
lookup table quadratically varying with the position in the image. The
PSF model is then refined by iteratively subtracting the neighbours of
the stars used to define the PSF, before computing the final iterated
PSF model.  A single PSF FWHM value is computed for each  `dither set', 
composed of the 4 detectors images, but a separate spatially-varying PSF 
model is calculated for each detector image. 
Finally the ALLSTAR procedure fits the model to all the stars
detected.  A second-pass detection is performed on the residual image
created by subtracting all sources detected in the first-pass.  This
second-pass uses a higher threshold of 8$- \sigma$, to avoid detecting
subtraction residuals.

DAOPHOT provides two measures of the reliability of a detection and of its photometric accuracy: the
goodness of fit, $\chi^2$, and a measure of the image sharpness, $sharp$, indicating image blemishes and resolved objects. 
Detections with $\chi^2>3$ and $\vert sharp \vert>2$ have been
eliminated. A more severe selection of stars with reliable
photometry is described below.  Detections around highly saturated
stars are manually deleted.

Standard stars from \cite{PER98} were observed to derive the magnitude
zero-point for each night. They were reduced as above, and analysed to
derive zero-point photometry using an aperture photometry radius of 20
pixels, equivalent to a diameter of 8 arcsec. Those observations show
that a large number of the fields of this survey where observed during
non-photometric nights. We thus require an external calibration in
these fields, which we derive from 2MASS.  The H and K$_s$ filters of
CIRSI and 2MASS are identical, ensuring a straightforward relative
calibration. However the J band filter of 2MASS is more extended into
the atmospheric water absorption features at around 1.1 and 1.4 $\mu$m
(\citealt{CAR01}), leading to a photometric zero-point shift of typically
J$_{(CIRSI)}-$J$_{(2MASS)} \sim -0.04$ mag.  To ensure that all our
photometry is on a homogeneous photometric system, we then calibrated
all our fields to the 2MASS catalogue zeropoints.  The calibration was
done on each dither set, the atmospheric conditions changing on this
time scale during non-photometric nights. By comparing this 2MASS
calibration to the standard star calibrations for good nights, the
zero-point calibration accuracy is better than 5\%.

\begin{figure}
\centering
\includegraphics[width=84mm]{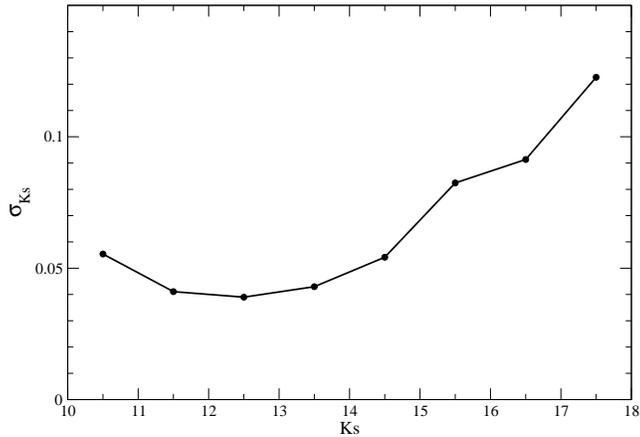}
\caption{Estimation of the photometric errors in the K$_s$ band from repeat observations
with a different central pointing in field 5PN.}   
\label{FsigK}
\end{figure}

To further test the accuracy of our data reduction and source extraction, 
we obtained a fifth full data set for field 5PN in the
K$_s$ band. This field was pointed at the centre of the standard data
sets for this field, and so provides an independent data set which
overlaps all the other dither sets. This independent set of photometry
was then analysed to determine the difference in the photometry for
this field as a function of magnitude. The resulting dispersion, 
presented in figure \ref{FsigK}, should be conservative as the 
overlapping areas are at the corner of the images where the PSF fitting is the 
least robust. On the other hand those dispersions could be larger during non-photometric nights. 
 
Combining the imaging data for the different filter passbands is 
straightforward.
Each single-colour image is divided into several small areas. In each
area for each colour, stars with very reliable photometry and astrometry
($\chi^2<1.5$ and $\vert sharp \vert<0.5$) are cross-correlated
between different passband images,
and matched with an allowed matching radius of 1.3 arcseconds. 
The mean shifts between the astrometric systems of the two passbands are then computed and applied to 
all stars. A final cross-identification is then defined, using these adjusted
astrometric solutions, with a smaller matching radius of 0.5 arcseconds. 
The JHK$_s$ catalogue is then derived from the JH and JK$_s$ catalogues, 
using the HK$_s$ cross-match to reject false and multiple cross-identifications. 

\begin{figure}
\centering
\includegraphics[width=84mm]{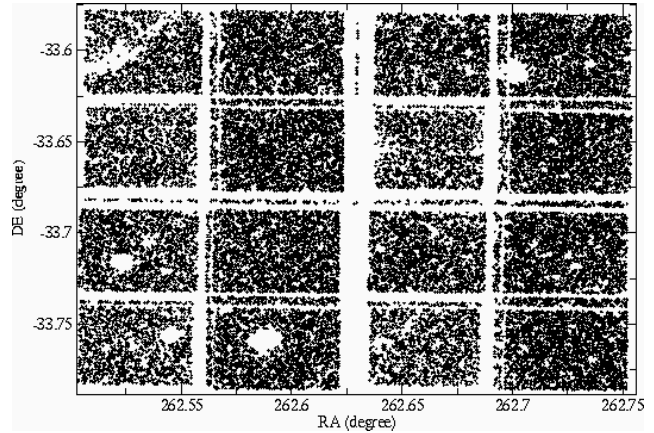}
\caption{Position of all stars detected in field 5NP in J H and K$_s$
 with $\chi^2<1.5$ and $\vert sharp \vert<0.5$.}   
\label{Fpos}
\end{figure}

Given the complexity of the source distributions in our fields, a
completeness estimation for number counts analyses would be unreliable. 
Indeed figure \ref{Fpos} illustrates the various instrumental, atmospheric 
and astrophysical effects on the number of stars detected. 
The gaps between the detectors are apparent due to our
removal of the dithered image borders. 
The photometric conditions influence the overall completeness.
 Variations of the number
counts between dither sets are due to changes in the atmospheric
conditions and in particular in the seeing.  Large holes are due to
the presence of saturated stars.  Variations of the number counts
within chip images are also visible due to variations of the
extinction. For example a dark cloud is clearly present in the
top-left corner of figure \ref{Fpos}.

\section{Red Clump Giants as Distance Indicators: Deriving the distances}

\begin{figure}
\centering
\includegraphics[width=84mm]{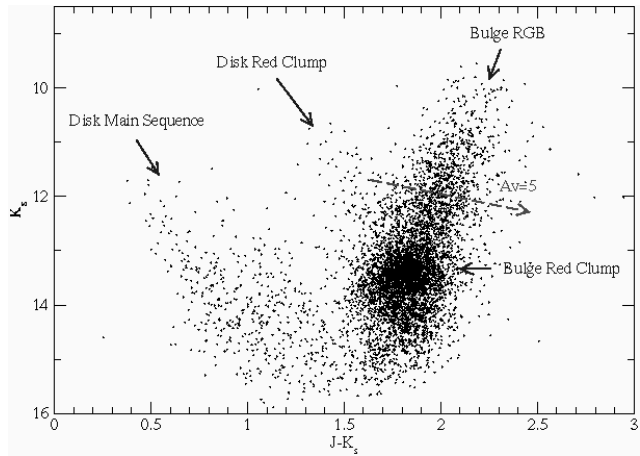}
\caption{Colour-magnitude diagram of the bulge field C32
($\ell=0\degr$, $b=1\degr$), using all stars detected in J,H and K$_s$ with 
$\chi^2<1.5$ and $\vert sharp \vert<0.5$.
The dotted arrow represents the movement of a star
on this diagram due to an absorption of $A_V$=5 mag.}
\label{CMDdesc}
\end{figure}

The main stellar populations observed in this survey are illustrated
in figure \ref{CMDdesc}, which is the colour-magnitude diagram (CMD) of the
minor axis low-extinction standard field C32.  In the blue part of the
diagram lie foreground main-sequence disc stars.  In field C32, since
the extinction is low and because the central inner Galactic bulge
clearly is spatially highly-concentrated with a small scale length,
the bulge red giant branch (RGB) stars form a well-defined, almost
single-distance RGB feature with a distinct red clump.  The foreground
disc red clump giants form a third sequence on the CMDs, roughly
parallel to the disk main sequence stars, becoming fainter and redder
as their distance and total extinction increase.  The narrow
intrinsic luminosity distribution of the red clump stars, apparent in this
figure, illustrates their utility as good distance indicators.

Given the high and variable extinction apparent in our fields,
reddening independent magnitudes are a critical requirement.  Such
parameters can be derived using any two colours, given an extinction
law, for example:
\begin{equation}
{K_s}_{J-K_s} = K_s - {A_{K_s} \over A_J-A_{K_s}}(J-K_s)
\label{eqme}
\end{equation}

In the near-infrared, the extinction curve is well defined by a single
power law function (e.g. \citealt{CAR89}, \citealt{MAT90},
\citealt{HE95}): $A_\lambda/A_J = (\lambda/\lambda_J)^{-\alpha}$.  As
a consequence, and unlike optical studies such as \cite{STA94}, where
extinction law variations remain a source of possible systematic
uncertainty, our reddening independent magnitudes should not be
significantly affected by variations of ISM properties along the
different lines of sight, although estimates of $A_V$ would.  In the
following we adopt the coefficient $\alpha=1.73$ of \cite{HE95},
leading to $A_{K_s}/E_{J-K_s}$ = 0.64, which agrees with the values
reviewed by \cite{MAT90}.

\begin{figure*}
\centering
\includegraphics[angle=-90,width=0.4\textwidth]{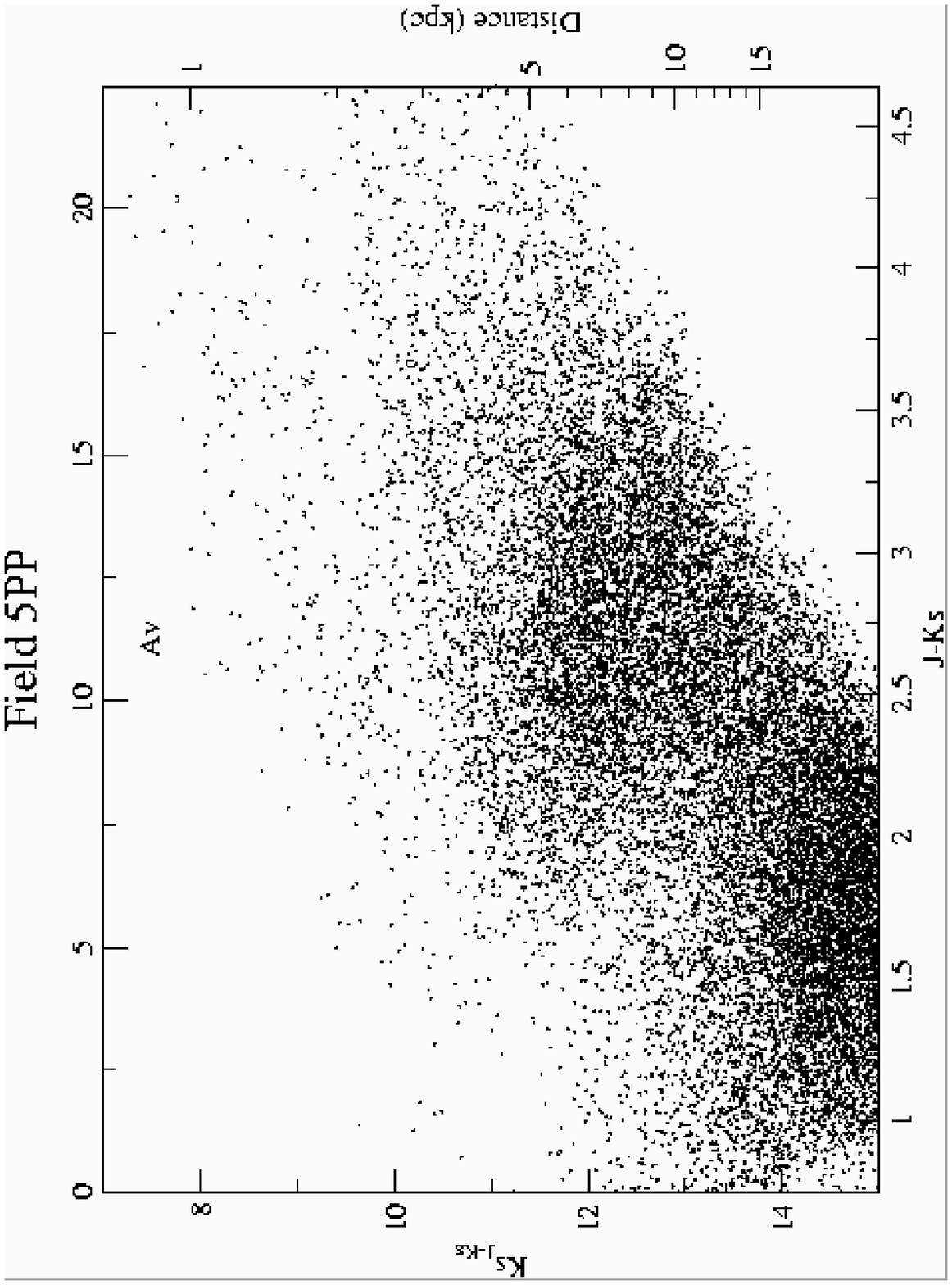}
\hspace{0.1\textwidth}
\includegraphics[angle=-90,width=0.4\textwidth]{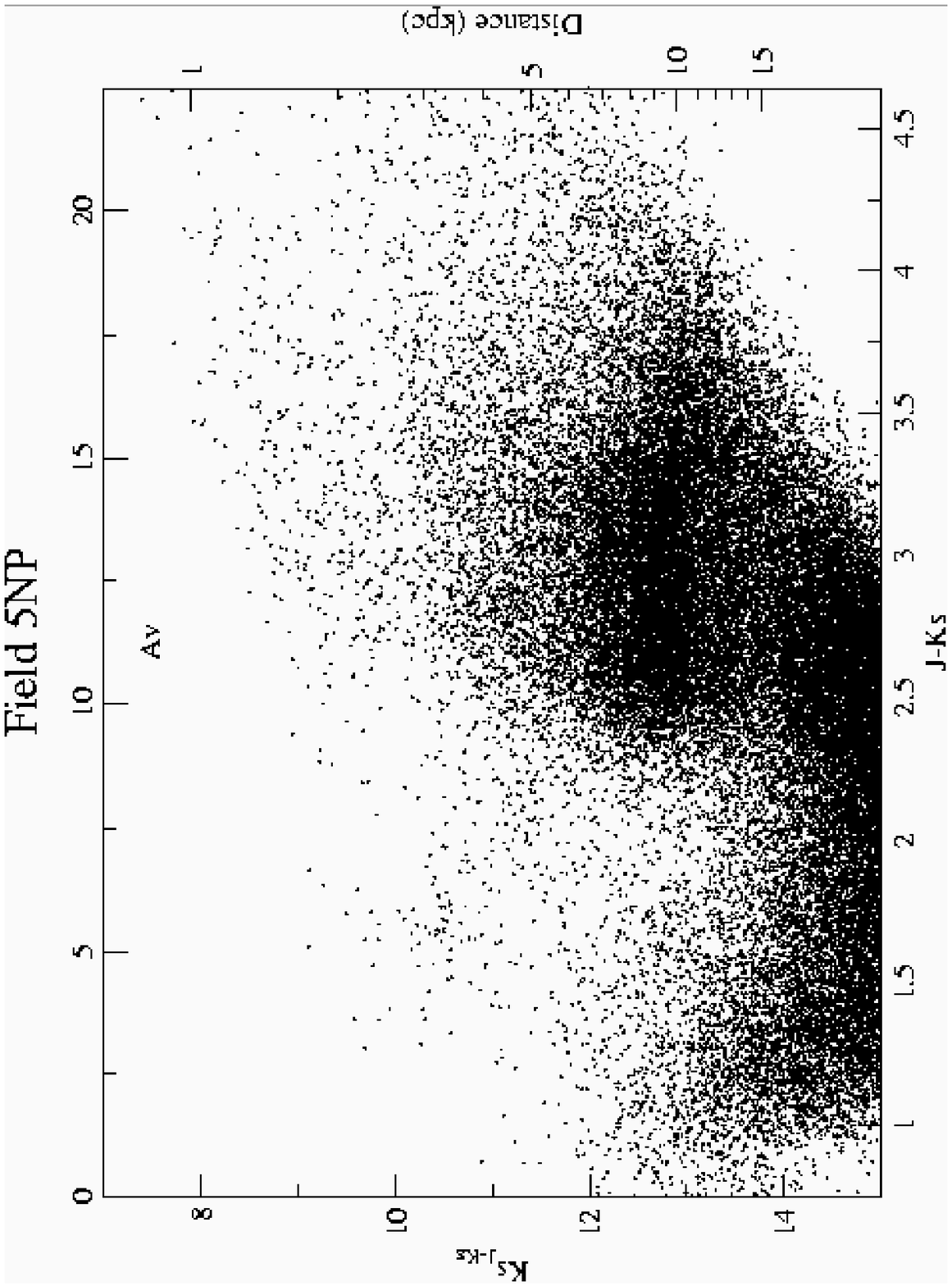}
\par \vspace{3mm}
\includegraphics[angle=-90,width=0.4\textwidth]{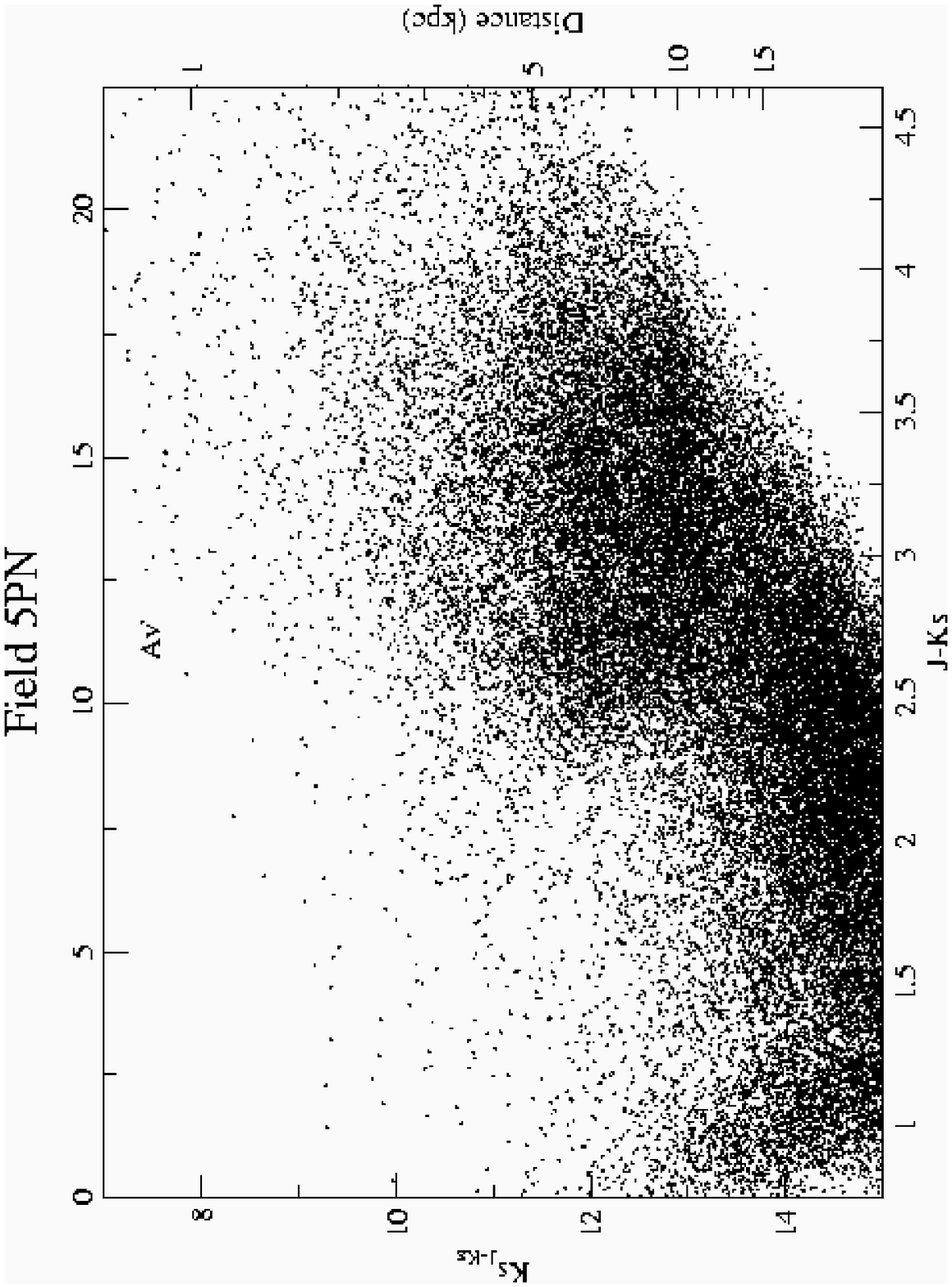}
\hspace{0.1\textwidth}
\includegraphics[angle=-90,width=0.4\textwidth]{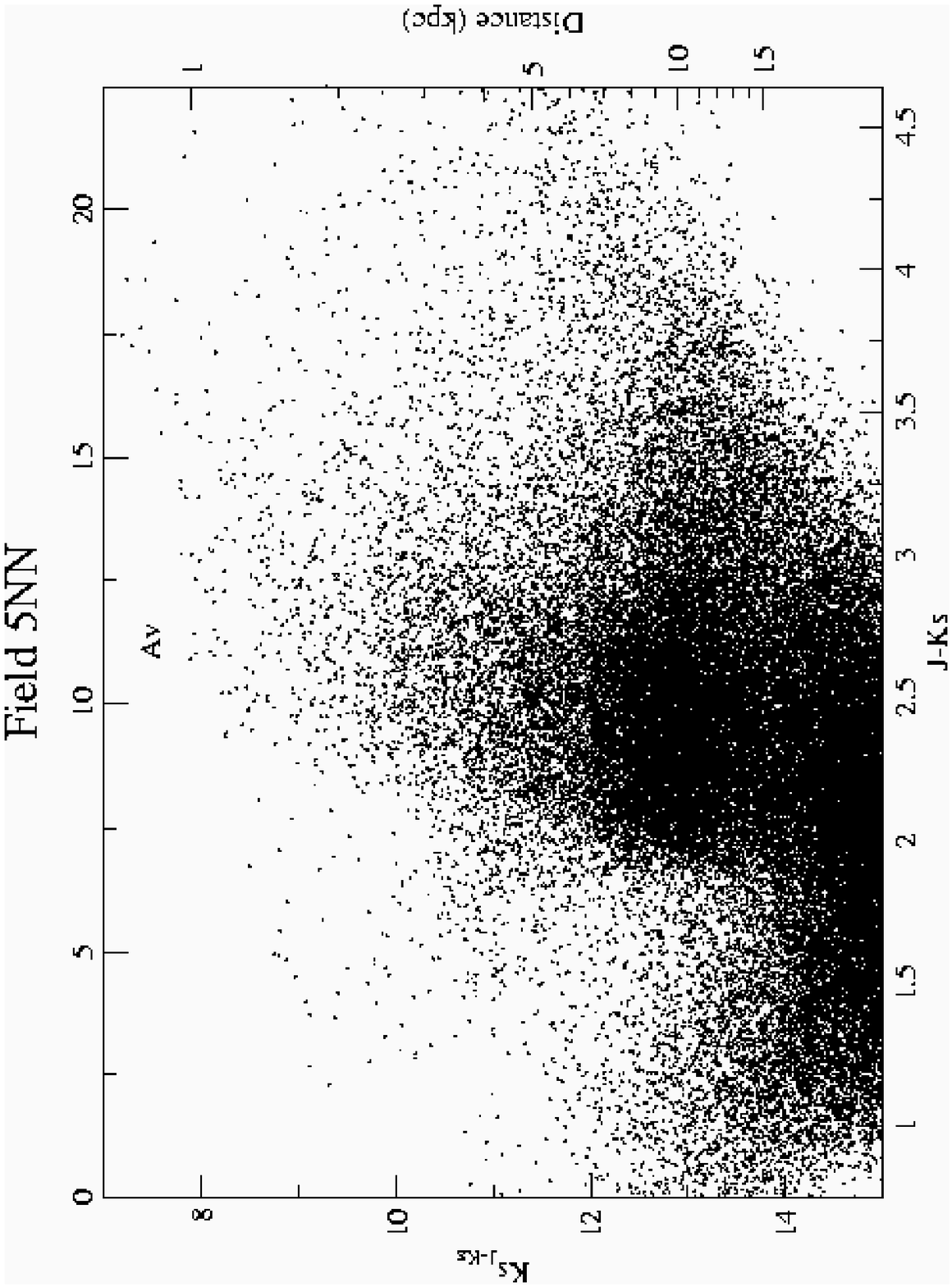}
\par \vspace{3mm}
\includegraphics[angle=-90,width=0.4\textwidth]{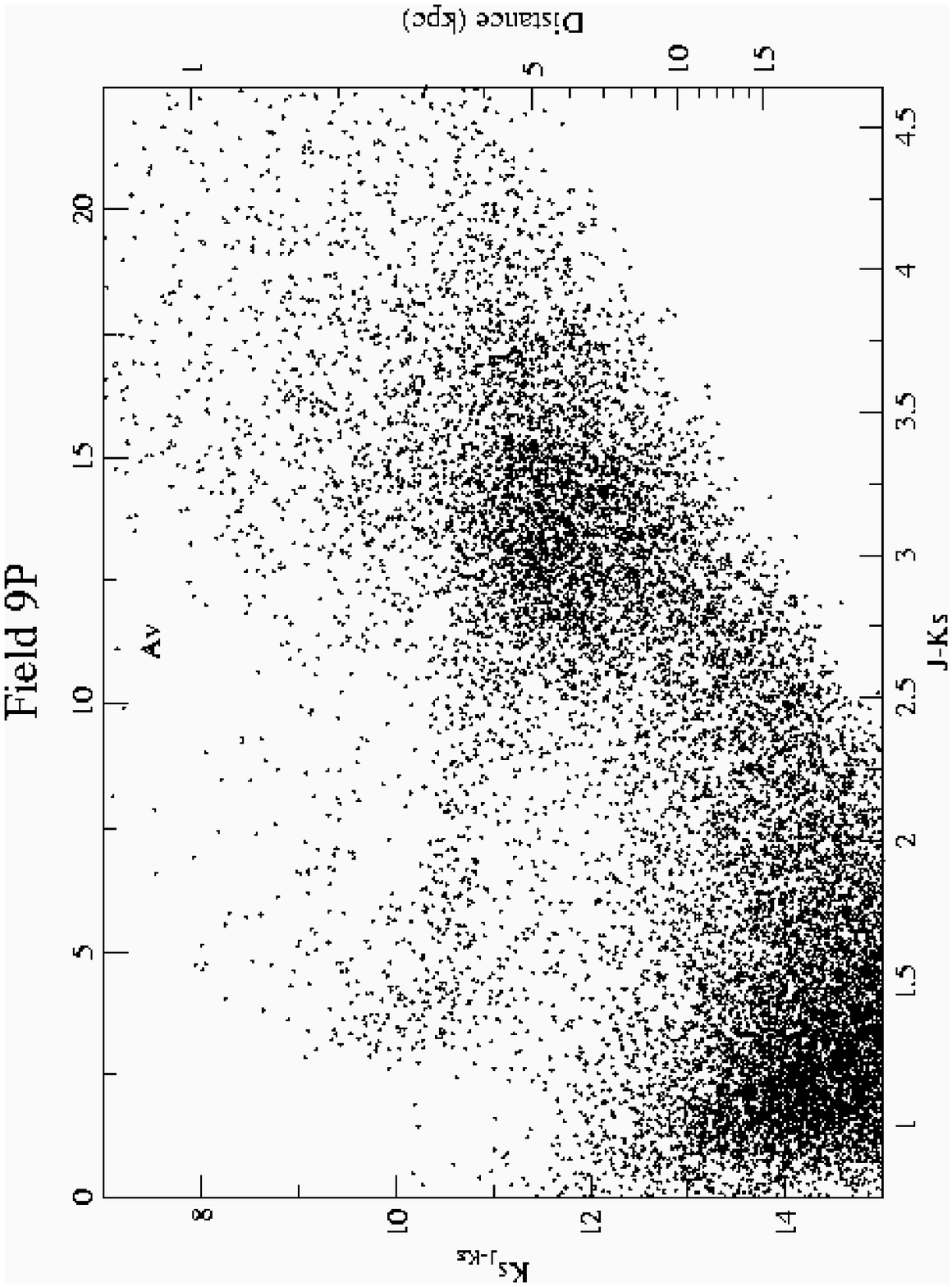}
\hspace{0.1\textwidth}
\includegraphics[angle=-90,width=0.4\textwidth]{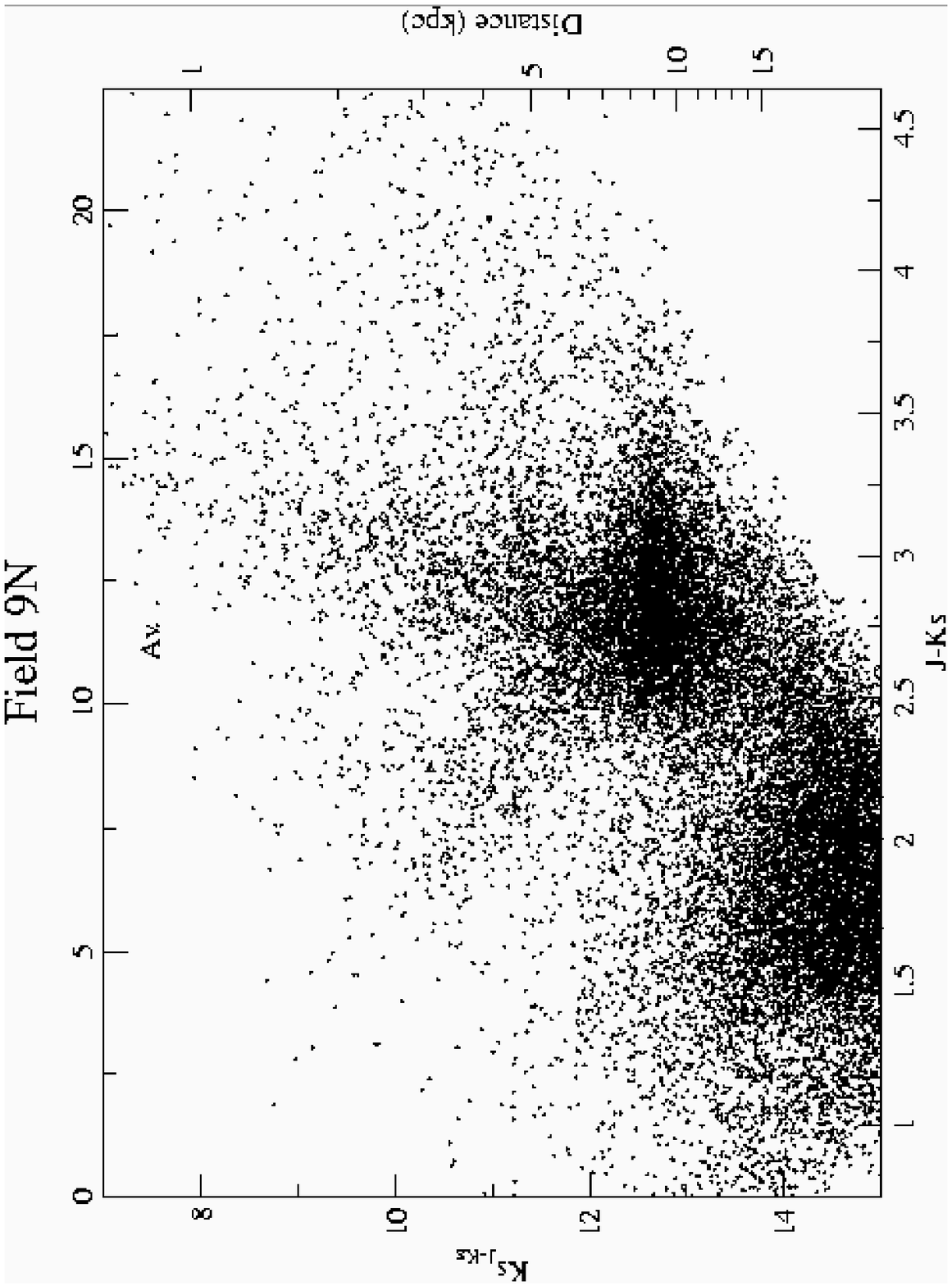}
\par \vspace{3mm}
\includegraphics[angle=-90,width=0.4\textwidth]{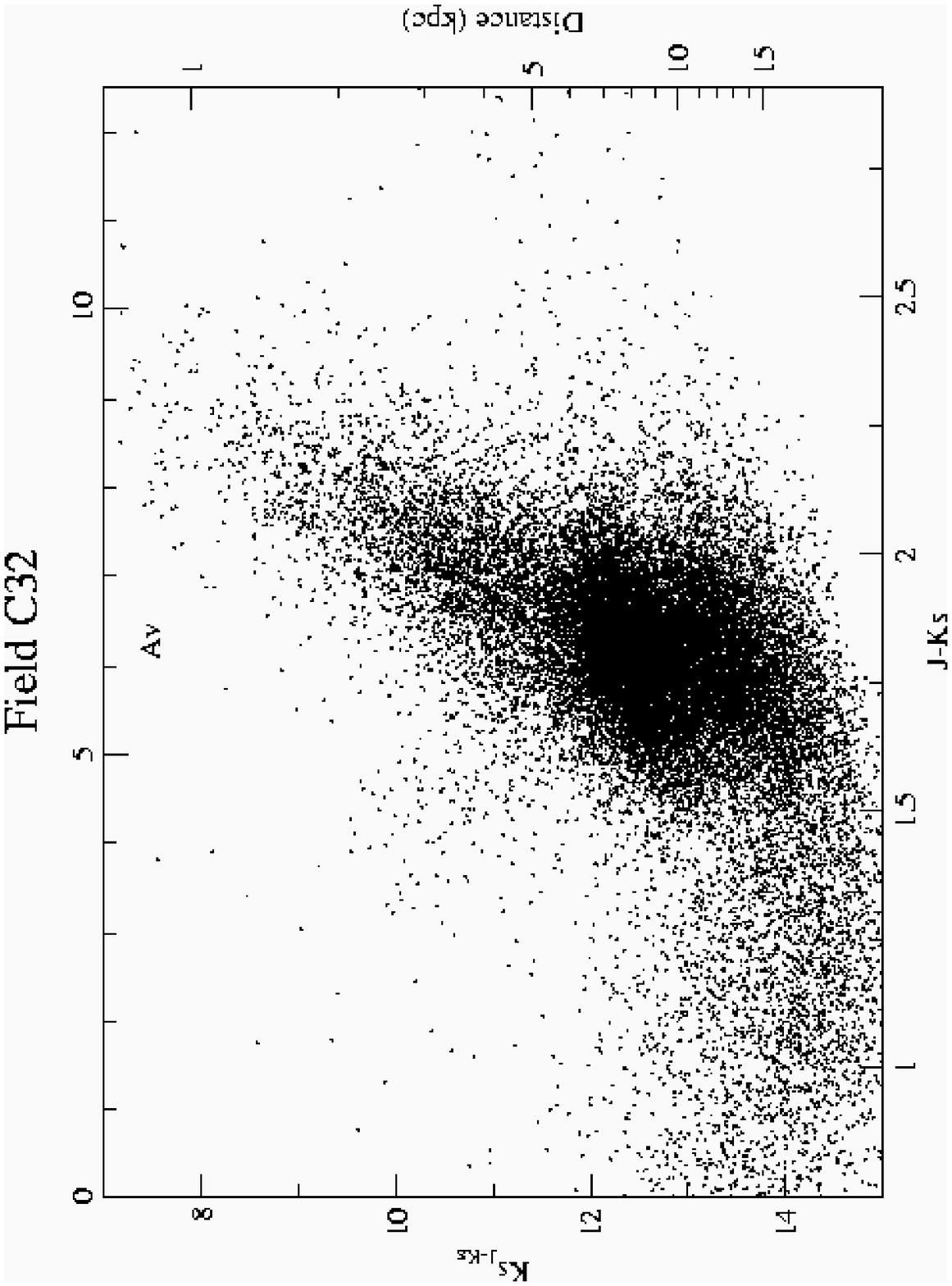}
\caption{Colour magnitude diagrams using reddening independent
magnitudes of all stars observed in J and K$_s$. 
In these figures, increasing extinction moves a star
horizontally to the right, while increasing distance moves it
vertically downwards. Locations of red clump stars on this diagram as
a function of absorption and distance are given by the top and right
hand axes.}
\label{mKJK}
\end{figure*}

In a colour magnitude diagram made with a reddening-independent
magnitude, the effect of distance and extinction are disentangled: a
star of a given spectral type moves vertically with distance,
horizontally with extinction.  Figure \ref{mKJK} illustrates, for each
of our seven observed fields, how the location of a red clump star in
the ($J-K_s$, ${K_s}_{J-K_s}$) diagram can be translated into an
($A_V$, distance) estimate. These diagrams therefore allow us to
quantify the distribution of both stellar density and of the
extinction along each line of sight.  For example, in field 9P a clear
flattening of the disc red clump distribution at ${K_s}_{J-K_s}
\approx 10.5$ indicates the presence of a specific source of
extinction located at about 3.5~kpc from the Sun, which increases
$A_V$ by about 7.5~mag in only 1.5~kpc. This feature is consistent
with the steep increase in absorption at 3.4-4~kpc seen at
$\ell$=12.9\degr\ by \cite{BER95}, that they associated with the
molecular ring.  In all our galactic plane fields, the bulk of bulge red 
clump stars become visible at $A_V \sim 10$ mag, and apparently
suffer a range of internal extinction of more than $\Delta
A_V\sim5$~mag.

A clear difference between the red clump distance 
distributions is present between the different lines of sight,
indicating that we have resolved the spatial structure of the inner Galaxy.

To quantify this, we determine the reddening-independent magnitude of the 
red clump, $m_{RC}$, by a non-linear least-squares fit of 
function \ref{eqfit} to the histogram of red giant stars 
(e.g. \citealt{STA98}, \citealt{SAL02}):
\begin{equation}
N(m) = a + b m + c m^2 + {N_{RC} \over \sigma_{RC} \sqrt{2\pi}}
\exp[-{(m_{RC}-m)^2 \over 2 \sigma_{RC}^2}]
\label{eqfit}  
\end{equation}
The Gaussian term represents a fit to the bulge/bar red clump.  The
first three terms describe a fit to the background distribution of
non-bulge-clump red giant stars, which in our case contains not only
the other bulge giants, but also the disc red clump stars that fall
into the bulge red clump.  The foreground dwarf star sequence is
eliminated on the CMDs by the criterion $K_s>4 (J-K_s) + b$, with b
determined for each CMD to account for the different extinction. As
distance and extinction grows, the mixing between faint dwarfs and
giants increases. Only stars with ${K_s}_{J-K_s}<14$ mag will then be
used, which corresponds to a limit in red clump star distance of about
15~kpc. These selection criteria ensure that all red clump stars at a
distance of 9~kpc with photometric error smaller than 3~$\sigma$ and
extinction smaller than $A_V$=15 are included in the analysis.

For given values of the red clump absolute magnitude and colour, a
reddening independent magnitude provides a direct estimate of the
distance modulus $\mu$ of red clump stars (equation \ref{distmod}),
and so a direct measure of the line-of-sight variation in Galactic
structure.

\begin{figure*}
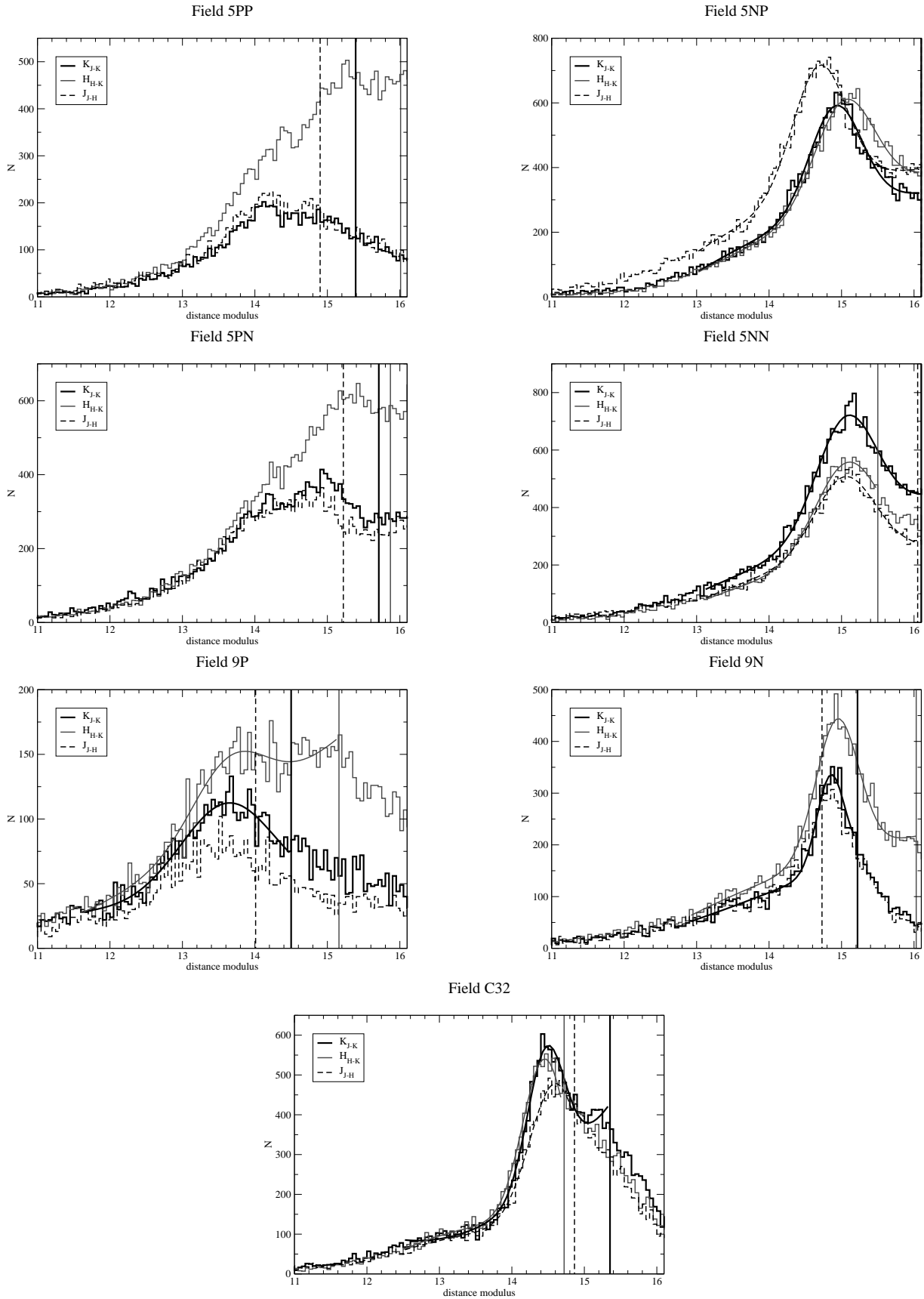

\centering
\includegraphics[width=0.4\textwidth]{fig5_5PP.eps}
\hspace{0.1\textwidth}
\includegraphics[width=0.4\textwidth]{fig5_5NP.eps}
\par \vspace{1.5mm}
\includegraphics[width=0.4\textwidth]{fig5_5PN.eps}
\hspace{0.1\textwidth}
\includegraphics[width=0.4\textwidth]{fig5_5NN.eps}
\par \vspace{1.5mm}
\includegraphics[width=0.4\textwidth]{fig5_9P.eps}
\hspace{0.1\textwidth}
\includegraphics[width=0.4\textwidth]{fig5_9N.eps}
\par \vspace{1.5mm}
\includegraphics[width=0.4\textwidth]{fig5_C32.eps}
\caption{The distance modulus distributions down each line of sight, computed for red clump stars using reddening-independent magnitudes. The vertical lines
show estimates of the photometric completeness limits. Fits to equation
\ref{eqfit} are overlaid, where available.} 
\label{histomod}
\end{figure*}

To check our calibration of the red clump colours, described below, 
and any effects of
photometric incompleteness, we have estimated the distance modulus of
the red clump from all three available reddening-independent
magnitudes: ${K_s}_{J-K_s}$, $H_{H-K_s}$ and $J_{J-H}$:
\begin{eqnarray}
\mu_{RC} &=&  {K_s}_{J-K_s} + A_K/E_{J-K} (J-K_s)_0 - {M_0}_{K_s} \nonumber \\
 &=&  H_{H-K_s} + A_H/E_{H-K} (H-K_s)_0 - {M_0}_{H} \nonumber \\
 &=&  J_{J-H} + A_J/E_{J-H} (J-H)_0 - {M_0}_{J} 
\label{distmod}
\end{eqnarray} 
These results are shown in figure \ref{histomod}.

We now require an upper limit to the reddening-independent magnitudes
that can be analysed reliably in the fit to equation \ref{eqfit}. For
this, we derive a rough estimate of the relative completeness of each
catalogue for each field ($J_C$,$H_C$,$K_C$) by determining when the
number counts as a function of magnitude stop increasing.  
The different atmospheric, instrumental and astrophysical effects 
influencing this relative completeness have been described in section 2. 
The relative completeness of each reddening independent magnitude is then
derived by combining these completeness estimates for each magnitude
and field, using the mean colour of the giants ($(J-K)_M$) and 
adding a margin of 0.5 mag to take into account the spread
in the mean colour:
\begin{eqnarray*}max(K_{J-K}) &=& min (J_C-(J-K)_M, K_C) \\
 && - A_K/E_{J-K} * (J-K)_M + margin
\end{eqnarray*}

The corresponding completeness-induced limits to our distance
determinations down each line of sight are indicated for each passband
and field in figure \ref{histomod} by vertical lines. A maximum value of 16.1 
is also apply due to the limit ${K_s}_{J-K_s}<14$ mag set before for the 
selection of giant stars. 

The model fits to the spatial distribution of bulge/bar red clump stars
down each line of sight are presented in figure \ref{histomod}. 
In general, the fits using the three different reddening independent
magnitudes are all consistent within about 0.1 mag, expect for field 5NP
where the dispersion is 0.2 mag. However no fit was possible using $J_{J-H}$ 
for fields 9P and 9N. 

The exception is the two fields at positive longitude $\ell = 5\degr$: no
fit converged for any of the distance modulus estimators for fields
5PP and 5PN. The large spread in the red clump colour and magnitudes
observed in those fields (figure \ref{mKJK}) cannot be explained as an
artefact of completeness nor by large photometric errors. The
interpretation of these results is discussed further in the next
section.

In order to convert our reddening-independent magnitudes into true distance
moduli, and hence into true spatial density distributions, we need to
calibrate the near-IR intrinsic luminosity of the red clump stars.
We used the true intrinsic colours of red clump stars predicted 
by the Padova isochrones in the 2MASS system (\citealt{BON04}) for a 10 Gyr 
old population of solar metallicity: 
($J-K_s$)$_0$=0.68 and ($H-K_s$)$_0$=0.07. 
We then calibrate the absolute magnitude of the red clump ${M_0}_{K_s}$ using
our data for field C32. The colour-magnitude
diagram for field C32, shown in figure \ref{CMDdesc}, is dominated by
the bulge and located at 0\degr\ in longitude, its red clump stars should 
therefore be at the distance of the Galactic centre. 
If we assume the absolute red clump distance of $M_K=-1.61 \pm 0.03$ mag, 
derived by \cite{ALV00} for the Hipparcos red clump, 
we obtain a distance for the Galactic center of $D_{GC}=7.6 \pm 0.15$ kpc. 
Comparing the Padova isochrones provided in the Johnson-Cousins-Glass 
system by \cite{GIR02} and the 2MASS ones, we find that no correction need 
to be applied between the \cite{ALV00} K photometry and our K$_s$ one for 
red clump stars. According to \cite{SAL02}, a small population correction 
should be applied to the \cite{ALV00} calibration for bulge stars. 
Using their population correction for Baade's window with solar metallicity
we derive $M_K=-1.68$ mag and $D_{GC}=7.8$ kpc, while
with enhanced $\alpha$-elements it leads to $M_K=-1.72$ mag 
and $D_{GC}=8.0$ kpc. However if we assume $\alpha$-enhancement, 
as we will see later on, we also have to change the assumed ($J-K_s$)$_0$, 
which leads us back to $D_{GC}=7.7 \pm 0.15$ kpc. 
Those values are consistent with the latest Galactic centre distance estimates
 (e.g. \citealt{REI93}, \citealt{MCN00}, \citealt{EIS03}) 
which give $D_{GC} = 8 \pm 0.5$ kpc.
Considering the different uncertainties in the absolute magnitude of the 
red clump, we decided to calibrate this latter to $M_{K_s}^{RC}=-1.72$ mag, 
assuming a distance for the Galactic centre of 8 kpc.

The results of our different fits to equation \ref{eqfit} have been
averaged and are presented in table \ref{distances}. The dispersion of the
Gaussian fitted to equation \ref{eqfit}, $\sigma_{RC}$, is the convolution
of the true line of sight dispersion in distance of the red clump stars ($\sigma_D$), the
intrinsic dispersion of the red clump luminosity ($\sigma_0$) and
photometric errors ($\sigma_e$). \cite{ALV00} estimates $\sigma_0$ to
be around 0.15-0.2 mag. The photometric errors are between 0.05 and
0.1 mag. An estimate of the dispersion due to the variation in
distance modulus is then indicated in table \ref{distances}, derived
from the simple deconvolution 
$\sigma_D = \sqrt{\sigma_{RC}^2 - \sigma_0^2 - \sigma_e^2}$.

\begin{table}
\caption{This table presents both the mean and the line-of-sight
dispersion in distances of bulge red clump stars for four different
galactic longitudes. $\sigma_m$ is the Gaussian dispersion of the fits
of equation \ref{eqfit} to the photometric data of figure \ref{mKJK},
which are shown in figure \ref{histomod}. $\sigma_D$ is our estimate of
the equivalent spread in distance, after deconvolution with the
intrinsic red clump luminosity dispersion and photometric errors.}
\label{distances}
\begin{center}
\begin{tabular}{cllll}
l & $\mu_{RC}$ (mag) & D (kpc) & $\sigma_{RC}$ (mag) & $\sigma_D$ (kpc) \\
\hline
-9.8\degr & 14.90$\pm$0.04 & 9.5$\pm$0.2 & 0.25$\pm$0.03 & 0.6 \\
-5.7\degr & 14.97$\pm$0.04 & 9.9$\pm$0.2 & 0.40$\pm$0.01 & 1.5 \\
0.0\degr  & 14.51$\pm$0.03 & 8           & 0.30$\pm$0.05 & 0.8 \\
+9.6\degr & 13.63$\pm$0.04 & 5.3$\pm$0.1 & 0.65$\pm$0.1  & 1.5 \\
\hline
\end{tabular}
\end{center}
\end{table}

We studied the robustness of our results against different choices of 
isochrones. A different age changes the absolute magnitude of the red clump 
but not its colour, so as we fix $M_{K_s}^{RC}$ from our data the choice of age does not 
affect our results. However metallicity and enrichment in $\alpha$-elements 
does affect the colours. A change in [Fe/H] of 0.4 dex changes $J-K_s$ 
by about 0.08 mag, the higher the metallicity the redder the colour. 
Isochrones for $\alpha$-enhanced stars lead to $J-K_s$ bluer by about 
0.1 mag. However as all is calibrated on field C32, the resulting distance 
estimates do not change by more than 0.3 kpc. However we note that the three 
different distance 
modulus estimates for field C32 (bottom of figure \ref{histomod}) agree much 
better with $\alpha$-enhanced isochrones (within 0.01 mag) than with the 
basic set (where the dispersion is 0.1 mag). 
On the other hand, for all the other fields the $\alpha$-enhanced isochrones lead to a larger dispersion. 
If we assume that we are not probing one but two different stellar populations, which would correspond to the bulge for field C32 and to a distinct Galactic bar for the other fields, such a difference would be expected.
Chosing the $\alpha$-enhanced colors for the bulge field (\citealt{MCW94}, \citealt{MAT99}) and the basic set for all the 
others would lead to an increase in distance modulus in all the `bar' fields 
of about 0.1 mag. If we also consider an age difference, using an age of 6 Gyr (\citealt{COL02})
for the bar fields would also add about 0.1 mag to the distance modulus. Figure \ref{PopsDiff} illustrates the effect of systematics in the assumed stellar populations: if we assume that the inner bulge is formed of an old population with enhanced $\alpha$ abundances, while the bar is formed of a younger population with solar abundances, the deduced distances show a much smaller dispersion from a galactic bar simple linear regression model. If confirmed by spectral observations, this implies a bar population which is formed from the inner disc, and not from the old bulge. 

\section{The structure of the inner Galaxy}

\begin{figure}
\centering
\includegraphics[width=0.45\textwidth]{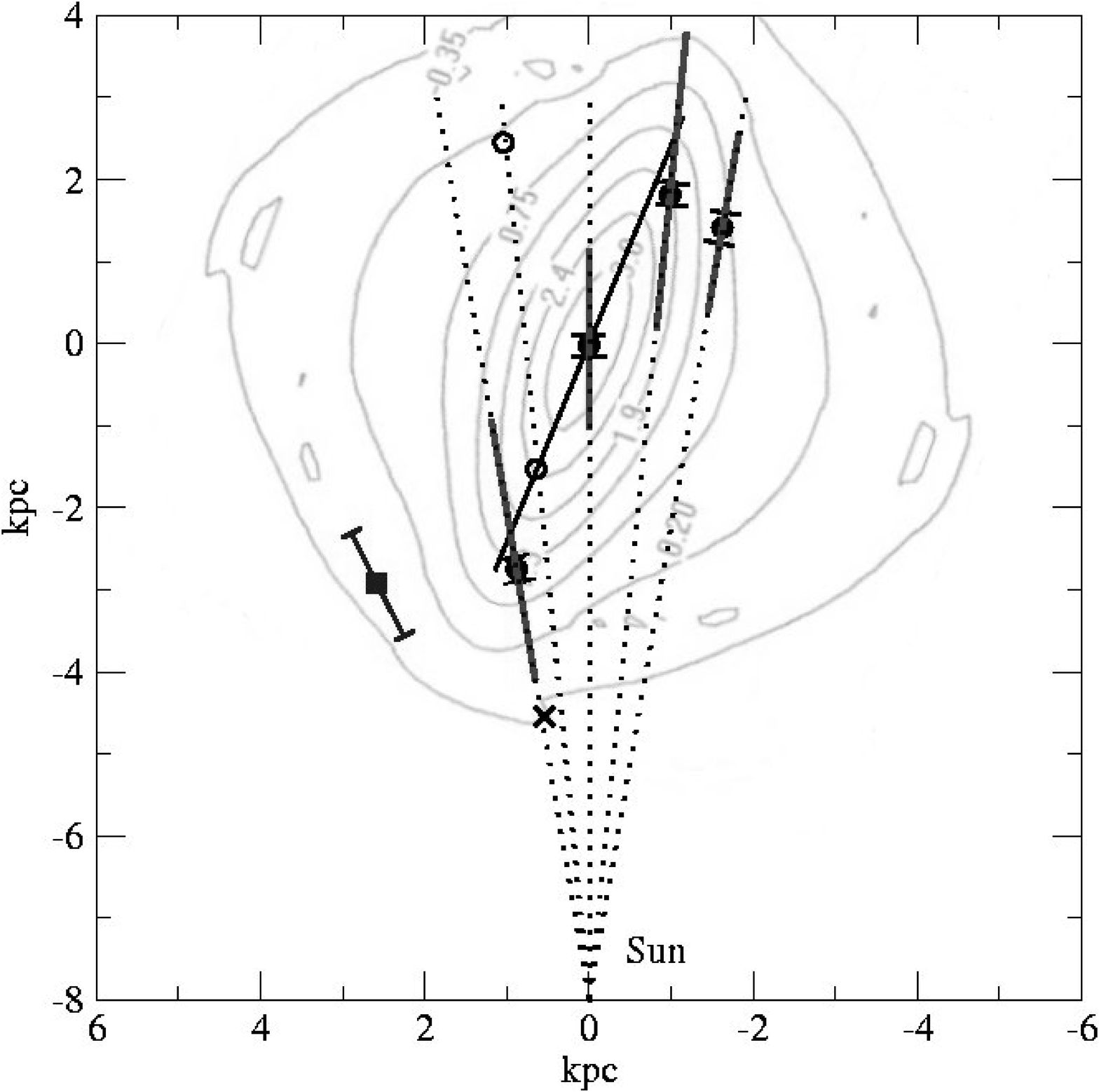}
\caption{The Galactic plane seen from the north Galactic pole.  Solid
circles indicate the mean positions of the red clump giants determined
in this analysis at Galactic longitudes of $\ell=\pm9.7\degr$,
$\ell=-5.7\degr$, and $\ell=0\degr$.  Thick grey lines along the lines of
sight through each mean distance represent the one$-\sigma$ range in
distances deduced from a Gaussian fit to the red clump apparent
distance modulus distribution, corrected for intrinsic red clump
luminosity dispersion and photometric errors.  The black line through
the mean distances illustrates a Galactic bar of 3~kpc radius inclined
at 22.5\degr\ to the Sun-Galactic Centre line.  The background contour
map is a plane projection of the Bissantz \& Gerhard (2002) galactic
bulge bar model, from their figure 11.  The empty circles at
$\ell=+5.7\degr$ are only indicative, as the more distant point may be
confused by background disk stars, as discussed in the text.  The small
cross at 3.5kpc from the Sun along the direction $\ell=+9.7\degr$
indicates the position of a local high-extinction region detected in
our photometry. The square to the left of the dotted lines through our
data indicates the mean position of the red clump detected by
Hammersley et al. (2000) at $\ell$=+27\degr.  }
\label{BarMod}
\end{figure}

\begin{figure}
\centering
\includegraphics[width=0.45\textwidth]{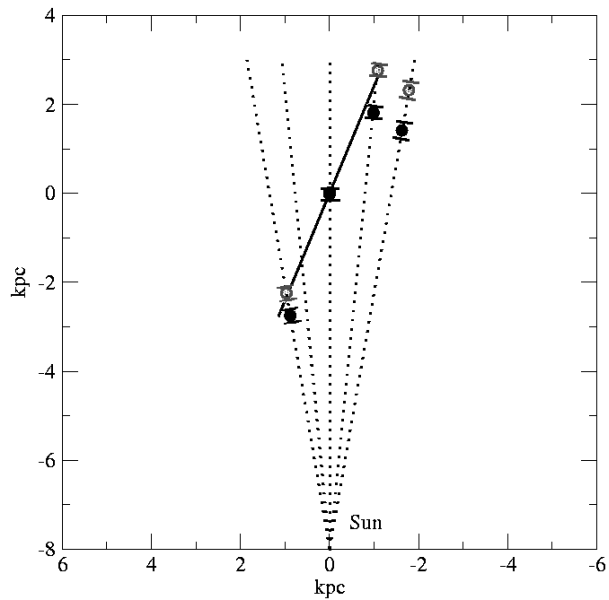}
\caption{Same as figure \ref{BarMod}, with the empty circles indicating the mean positions of the red clump giants determined with different isochrones for the bulge field at $\ell=0\degr$ and the other fields. 
The solid circles distances assume that the stellar population of all the fields are 10 Gyr old and of solar metallicity.
The empty circles positions are computed assuming for the bulge field at $\ell=0\degr$ a 10 Gyr old population with enhanced $\alpha$-element abundances, while assuming for all the other fields a 6 Gyr old population of solar abundances.
}
\label{PopsDiff}
\end{figure}

Figure \ref{BarMod} summarises our derived spatial map of the inner
Galaxy.  As well as the mean distances down each line of sight, the
range of distances corresponding to the deconvolved one-$\sigma$
dispersion in distance modulus is also illustrated. In spite of this
representation, we emphasise that the true distribution in distance
has no reason to be Gaussian, as it is linked to the geometry of the
bar/bulge and to our viewing angle. Our observation angle biases both
our distance and dispersion estimates, because of the larger volume,
and hence probable larger number of stars present in the far side of
our observation cone than in the near side. One should then keep in
mind that our distance estimates will be slightly biased towards
larger distances, especially for structures with a large true
dispersion in distances, and as close to us as is observed at
$\ell$=+9\degr.

Our mean red clump distances measured at $\ell=+9\degr$ and
$\ell=-5\degr$ are consistent with the more recent bar models,
particularly those derived considering both gas dynamics and
available surface brightness luminosity distributions. Those models
deduce a bar orientation with respect to the Sun -galactic centre
direction of 15\degr $\la \phi_{bar} \la $ 35\degr
(e.g. \citealt{GER01}, \citealt{MER03}). The model of \cite{BIS02}
(their figure 11) is overlaid on figure \ref{BarMod}. This model
assumes a bar with an angle $\phi_{bar}=20$\degr, a length of 3.5~kpc
and an axis ratio 10:3-4. The overall agreement with our direct
distance determinations is rather good.

Our mean red clump distance derived at $\ell=-9\degr$ is however not
simply consistent with the bar model described above. A single linear
fit to all the four mean distances measured at $\ell=+9,0,-5,+9\degr$
has all four points more than 3$\sigma$ from the `best fit'. However
the feature observed at $\ell=-9\degr$ does seem real: inspection of
figure \ref{mKJK} shows a clear overdensity of the whole red giant
branch at the same distance as the mean clump distance, determined to
be $\sim$ 9~kpc, in this field. Moreover, in this line of sight, the
most likely source of distance bias, the presence of disc red clump
stars, would bias the distance estimate towards larger distances, not
shorter. The red clump distance estimate at $\ell=-9\degr$ is
consistent with that at $\ell=-5\degr$. A possible physical
interpretation is that we have detected the signature of the end of
the bar. Star count peaks have also been detected in this region in
studies of old OH/IR stars by \cite{SEV99} and in the DENIS star
counts by \cite{LOP01}. \cite{LOP01} suggest that this local density
maximum may be related to the red clump overdensity detected by
\cite{HAM00} at $\ell=27\degr$. The \cite{HAM00} density maximum is
indicated in figure \ref{BarMod} by a square.

The presence of these two local features at $\ell=27\degr$ and $\ell=-9\degr$ 
and their mean red clump distance estimates
could imply the presence of a double triaxiality in the inner Galaxy, 
a double bar, or a triaxial bulge oriented at $\sim22\degr$ with a longer, 
thiner bar oriented at $\sim44\degr$. 
However, if such a second structure did exist, we should have detected 
its signature in our other survey fields. No such complex signature is evident
(cf. figure \ref{mKJK}). If two complex spatial distributions were projected 
down our lines of sight without being resolved, this would imply that our 
distance determinations are biased so that 
the angle of the first structure is in fact smaller than the 22\degr\ we
measured, and that the spread of the red clump distances observed down
each line of sight would encompass both structures. The latter does
not seem to correspond to the rather small distance dispersions we
derived, which are summarised in table \ref{distances}. Furthermore,
\cite{PIC03} do not detect a density excess at l=+21\degr, confirming
that the structure seen at l=+27$\degr$ is probably local, and
unrelated to the larger bulge or bar.

An alternative, and perhaps more consistent, interpretation of the
$\ell=-9\degr$ structure we observe is the
presence of a stellar ring or pseudo-ring at the end of the Galactic
bar. If so, the bar and the stellar ring would have a radius of
2.3$\pm$0.25~kpc. Our observations  at $\ell=+9\degr$ indicate that 
the radius of
the bar is at least 2.7$\pm$0.2~kpc long. Those two determinations are
consistent within one$-\sigma$. We note that assuming the bulge/bar stellar 
population differences used for figure \ref{PopsDiff} leads to a larger bar 
radius of about 3 kpc.
A bar radius of about 2.5~kpc would agree
with the model of \cite{LEP00} and \cite{SEV99}, but is smaller than
the value indicated by \cite{GER01} in his review.  The presence of
a ring has been suspected by several authors.  A molecular ring at
about 4-5~kpc is a well known feature deduced from CO maps. Our
detection of a step increase in the extinction distribution in field
9P (indicated by a cross in figure \ref{BarMod}) could be associated
with this molecular ring. \cite{COM96} link their derived distribution of
ultracompact HII regions to the molecular ring, but also detect a star
forming ring at about 2~kpc from the galactic centre.  A stellar ring
was suspected to lie at 3.5~kpc from the galactic centre by
\cite{BER95}.  From an OH/IR star study \cite{SEV99} suggests that an
inner ring lies between between 2.2 and 3.5~kpc. From star counts in
the DENIS survey \cite{LOP01} also argue for the presence of a stellar
ring, mainly from the detection of a density peak at $\ell=-22\degr$ that,
following \cite{SEV99}, they associate with the tangential point to
the 3-kpc arm, which is likely to be a (pseudo-)ring. The presence of
a Galactic ring would also help to reproduce the observed microlensing
optical depth (\citealt{SEV01}).  

Inner ring structures are indeed frequently observed in barred
spirals. Most inner rings are in fact pseudo-rings, formed from the
complex merging of inner bars and spiral structure
(e.g. \citealt{BUT96}). Such a pseudo-ring structure is consistent
with all the various detections of sub-structures between 2 and 4~kpc
from the Galactic centre.

\cite{LOP01} detect an asymmetry in the Galactic plane extinction distribution,
 with more extinction at negative than at positive longitudes. 
We do not confirm this result. The fact that we do not find any
asymmetry in red clump giant extinction between positive and negative
longitudes (figure \ref{mKJK}), while there is an asymmetry in
distance, is certainly consistent with a minimum in the dust
distribution in this region. 
Interior of the bar outer radius a lower global 
density of stars and gas is often seen in observations of other galaxies.  
Several authors have suggested that such a stellar density decrease may 
be present in the Galactic disc (\citealt{BIN91}, \citealt{BER95}, 
\citealt{FRE98}, \citealt{LEP00}, \citealt{LOP04}). Our observations are 
consistent with such a decrease in the dust distribution of the inner disc. 

 Another interesting feature often observed in barred galaxies is the 
presence of dust lanes associated with the bar. From DIRBE surface 
brightness maps, \cite{CAL96} suggested the presence of such a dust lane, 
preceding the Galactic bar. They estimate the extra absorption present at 
negative longitudes to be between 1 and 2.6 mag in K. The presence of such 
a feature is clearly ruled out by our observations.

\subsection{The line of sight towards $\ell=+5\degr$}

\begin{figure}
\centering
\includegraphics[width=84mm]{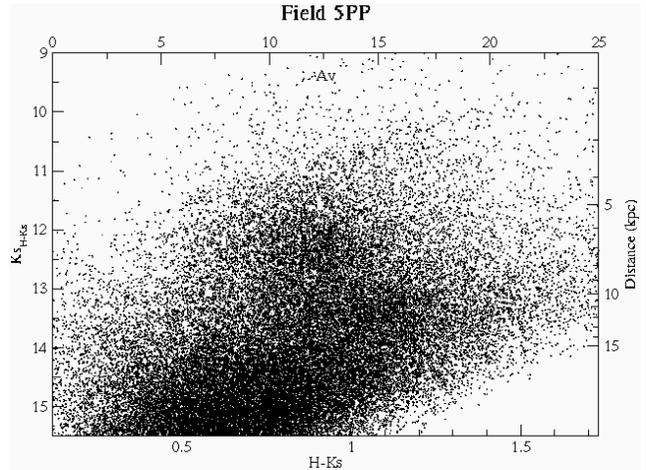}
\caption{Colour - reddening independent magnitude diagram for field 5PP
($\ell=+5\degr$). This figure is derived from H\&K$_s$ photometry,
which have the faintest completeness limits for this field.}
\label{KeHK}
\end{figure}

The reddening-independent colour-magnitude diagram for the lines of
sight towards $\ell=+5\degr$ (figure \ref{KeHK}) differ significantly
from those in other directions, in that the red clump stars indicate a
very considerable range of distances, with no clear local maximum
density. This is also very apparent in the relevant panels of figure
\ref{histomod}, which show the corresponding distribution in distance
modulus.  The broad apparent distance distribution observed for the
red clump stars at $\ell=+5\degr$ could be explained by the presence
of two features, one at $\sim$ 6~kpc and a second at $\sim$
11~kpc. Indeed, careful inspection of figure \ref{KeHK} suggests that
the clump at ${K_s}_{H-K_s}\sim12.5$ does not extend redder than
H-K$_s$$>$1.2, while the fainter part, corresponding to distances
greater than $\sim10$kpc, extends beyond the photometric completeness
limit. The first distance range is consistent with the bar structure
confirmed above. It is also consistent with the fact that the OGLE
data of \cite{STA94}, observed at a lower latitude of b=$-3.5\degr$, 
show similar
photometric behaviour at $\ell=+5\degr$ compared to $\ell=-5\degr$,
except for the expected shift in distance modulus. 

The fainter feature in our low-latitude data could then be due to
another structure, further away. This feature could be due to disc red
clump stars. Indeed with increasing distance both the volume observed
and the extinction increase, and considering the logarithmic relation
between distance and magnitude, any distant red clump stars visible
would concentrate in the high-extinction high distance-modulus
bottom-right part of the CMD. Deeper data than we have available are
then needed to determine if we are indeed seeing the distant disk
beyond the bulge or a local structure which could be associated with the ring. 
This region is also interesting 
for further study as a broad velocity-width molecular clump has been 
detected at
$5.2\degr<\ell<6.0\degr$ (e.g. \citealt{BOY94}, \citealt{BIT97},
\citealt{DAM01}), which may be a manifestation of gas shocks
(\citealt{KUM97}, \citealt{FUX99}). An estimate of the distance of
those two structures, if the fainter one is real and indeed a local
maximum, is presented in figure \ref{BarMod} as the open circles.

\section{Conclusion}

We have obtained, reduced and analysed deep wide-field near-infrared
photometry of five lines of sight towards the Galactic bulge within
0.25\degr\ of the Galactic plane, using the CIRSI camera. This has
provided a substantial improvement in our quantitative knowledge of
inner Galactic structure, and in particular of the structure of the
Galactic bar. Our use of near-infrared data allows us to determine
reddening-independent magnitudes and to observe bulge red clump stars
within the galactic plane, where the extinction is too high for
optical studies. This allows the detection of tracer stellar
populations independently of their scale height.  Colour -
reddening-independent magnitude diagrams have been shown to
disentangle the effects of distance and extinction, allowing a direct
conversion of red-clump star photometry ($J-K$,$K_{J-K}$) into an
($A_V$,distance) estimate.  Those determinations have led to the
following main results:
\begin{enumerate}
\renewcommand{\theenumi}{-}
\item The presence of a triaxial structure at the centre of our Galaxy
is confirmed.  Its angle relative to the Sun-Galactic centre line is
$\phi_{bar}=22 \pm 5.5 $\degr. It extends to at least 2.5~kpc from
the Galactic Centre. A large axis ratio is excluded, but our data are consistent
with a 10:3-4 ratio. In particular the distance dispersion of the bulge along
the ($\ell=$0\degr,b=1\degr) line of sight is less than 1~kpc.
\item A structure present at $\ell=-9.8\degr$ is not aligned with this
triaxiality. We suggest that the structure present at $\ell=-9.8\degr$ is
likely to be the signature of the end of the Galactic bar, which is
therefore circumscribed by an inner pseudo-ring.
\item A decrease in the dust distribution inside the bar radius is
inferred from the extinction distribution in our fields.
\item Our observations are not consistent with the existence of the
dust lane preceding the Galactic bar at negative longitudes suggested
by \cite{CAL96}.
\end{enumerate}
  
\section*{Acknowledgments}
We are grateful to Jacco van Loon and Robert Sharp for their
participation in the CIRSI observations, and to the CIRSI team for
building the camera.  The development and construction of CIRSI was
made possible by a generous grant from the Raymond and Beverly Sackler
Foundation.  This publication makes use of data products from the Two
Micron All Sky Survey, which is a joint project of the University of
Massachusetts and the Infrared Processing and Analysis
Center/California Institute of Technology, funded by the National
Aeronautics and Space Administration and the National Science
Foundation.

\bibliographystyle{mn2e}

\begin{thebibliography}{}

\bibitem[\protect\citeauthoryear{{Alves}}{{Alves}}{2000}]{ALV00}
{Alves} D.~R.,  2000, \apj, 539, 732

\bibitem[\protect\citeauthoryear{{Beckett}, {Mackay}, {McMahon}, {Parry},
  {Piche} \& {Ellis}}{{Beckett} et~al.}{1997}]{BEC97}
{Beckett} M.~G.,  {Mackay} C.~D.,  {McMahon} R.~G.,  {Parry} I.~R.,  {Piche}
  F.,    {Ellis} R.~S.,  1997, in Proc. SPIE Vol. 2871, p. 1152-1159, Optical
  Telescopes of Today and Tomorrow, Arne L. Ardeberg; Ed. {CIRSI: the Cambridge
  Infrared Survey Instrument}.
pp 1152--1159

\bibitem[\protect\citeauthoryear{{Bertelli}, {Bressan}, {Chiosi}, {Ng} \&
  {Ortolani}}{{Bertelli} et~al.}{1995}]{BER95}
{Bertelli} G.,  {Bressan} A.,  {Chiosi} C.,  {Ng} Y.~K.,    {Ortolani} S.,
  1995, \aap, 301, 381

\bibitem[\protect\citeauthoryear{{Bertin} \& {Arnouts}}{{Bertin} \&
  {Arnouts}}{1996}]{BER96}
{Bertin} E.,  {Arnouts} S.,  1996, A\&A Suppl. Ser., 117, 393+

\bibitem[\protect\citeauthoryear{{Binney}, {Bissantz} \& {Gerhard}}{{Binney}
  et~al.}{2000}]{BIN00}
{Binney} J.,  {Bissantz} N.,    {Gerhard} O.,  2000, \apjl, 537, L99

\bibitem[\protect\citeauthoryear{{Binney}, {Gerhard}, {Stark}, {Bally} \&
  {Uchida}}{{Binney} et~al.}{1991}]{BIN91}
{Binney} J.,  {Gerhard} O.~E.,  {Stark} A.~A.,  {Bally} J.,    {Uchida} K.~I.,
  1991, \mnras, 252, 210

\bibitem[\protect\citeauthoryear{{Bissantz} \& {Gerhard}}{{Bissantz} \&
  {Gerhard}}{2002}]{BIS02}
{Bissantz} N.,  {Gerhard} O.,  2002, \mnras, 330, 591

\bibitem[\protect\citeauthoryear{{Bitran}, {Alvarez}, {Bronfman}, {May} \&
  {Thaddeus}}{{Bitran} et~al.}{1997}]{BIT97}
{Bitran} M.,  {Alvarez} H.,  {Bronfman} L.,  {May} J.,    {Thaddeus} P.,  1997,
  \aaps, 125, 99

\bibitem[\protect\citeauthoryear{{Blitz} \& {Spergel}}{{Blitz} \&
  {Spergel}}{1991}]{BLI91}
{Blitz} L.,  {Spergel} D.~N.,  1991, \apj, 379, 631

\bibitem[\protect\citeauthoryear{{Bonatto}, {Bica} \& {Girardi}}{{Bonatto}
  et~al.}{2004}]{BON04}
{Bonatto} C.,  {Bica} E.,    {Girardi} L.,  2004, \aap, 415, 571

\bibitem[\protect\citeauthoryear{{Boyce} \& {Cohen}}{{Boyce} \&
  {Cohen}}{1994}]{BOY94}
{Boyce} P.~J.,  {Cohen} R.~J.,  1994, \aaps, 107, 563

\bibitem[\protect\citeauthoryear{{Buta} \& {Combes}}{{Buta} \&
  {Combes}}{1996}]{BUT96}
{Buta} R.,  {Combes} F.,  1996, Fundamentals of Cosmic Physics, 17, 95

\bibitem[\protect\citeauthoryear{{Calbet}, {Mahoney}, {Hammersley}, {Garzon} \&
  {Lopez-Corredoira}}{{Calbet} et~al.}{1996}]{CAL96}
{Calbet} X.,  {Mahoney} T.,  {Hammersley} P.~L.,  {Garzon} F.,
  {Lopez-Corredoira} M.,  1996, \apjl, 457, L27+

\bibitem[\protect\citeauthoryear{{Cardelli}, {Clayton} \& {Mathis}}{{Cardelli}
  et~al.}{1989}]{CAR89}
{Cardelli} J.~A.,  {Clayton} G.~C.,    {Mathis} J.~S.,  1989, \apj, 345, 245

\bibitem[\protect\citeauthoryear{{Carpenter}}{{Carpenter}}{2001}]{CAR01}
{Carpenter} J.,  2001, AJ, 121, 2851+

\bibitem[\protect\citeauthoryear{{Cole} \& {Weinberg}}{{Cole} \&
  {Weinberg}}{2002}]{COL02}
{Cole} A.~A.,  {Weinberg} M.~D.,  2002, \apjl, 574, L43

\bibitem[\protect\citeauthoryear{{Comeron} \& {Torra}}{{Comeron} \&
  {Torra}}{1996}]{COM96}
{Comeron} F.,  {Torra} J.,  1996, \aap, 314, 776

\bibitem[\protect\citeauthoryear{{Dame}, {Hartmann} \& {Thaddeus}}{{Dame}
  et~al.}{2001}]{DAM01}
{Dame} T.~M.,  {Hartmann} D.,    {Thaddeus} P.,  2001, \apj, 547, 792

\bibitem[\protect\citeauthoryear{{de Vaucouleurs}}{{de
  Vaucouleurs}}{1964}]{DEV64}
{de Vaucouleurs} G.,  1964, in IAU Symp. 20: The Galaxy and the Magellanic
  Clouds {Interpretation of velocity distribution of the inner regions of the
  Galaxy}.
pp 195--+

\bibitem[\protect\citeauthoryear{{Deguchi}, {Fujii}, {Izumiura}, {Kameya},
  {Nakada} \& {Nakashima}}{{Deguchi} et~al.}{2000}]{DEG00}
{Deguchi} S.,  {Fujii} T.,  {Izumiura} H.,  {Kameya} O.,  {Nakada} Y.,
  {Nakashima} J.,  2000, \apjs, 130, 351

\bibitem[\protect\citeauthoryear{{Dwek}, {Arendt}, {Hauser}, {Kelsall},
  {Lisse}, {Moseley}, {Silverberg}, {Sodroski} \& {Weiland}}{{Dwek}
  et~al.}{1995}]{DWE95}
{Dwek} E.,  {Arendt} R.~G.,  {Hauser} M.~G.,  {Kelsall} T.,  {Lisse} C.~M.,
  {Moseley} S.~H.,  {Silverberg} R.~F.,  {Sodroski} T.~J.,    {Weiland} J.~L.,
  1995, \apj, 445, 716

\bibitem[\protect\citeauthoryear{{Eisenhauer}, {Sch{\" o}del}, {Genzel}, {Ott},
  {Tecza}, {Abuter}, {Eckart} \& {Alexander}}{{Eisenhauer}
  et~al.}{2003}]{EIS03}
{Eisenhauer} F.,  {Sch{\" o}del} R.,  {Genzel} R.,  {Ott} T.,  {Tecza} M.,
  {Abuter} R.,  {Eckart} A.,    {Alexander} T.,  2003, \apjl, 597, L121

\bibitem[\protect\citeauthoryear{{Freudenreich}}{{Freudenreich}}{1998}]{FRE98}
{Freudenreich} H.~T.,  1998, \apj, 492, 495

\bibitem[\protect\citeauthoryear{{Fux}}{{Fux}}{1999}]{FUX99}
{Fux} R.,  1999, \aap, 345, 787

\bibitem[\protect\citeauthoryear{{Gerhard}}{{Gerhard}}{2001}]{GER01}
{Gerhard} O.~E.,  2001, in ASP Conf. Ser. 230: Galaxy Disks and Disk Galaxies
  {Structure and Mass Distribution of the Milky Way Bulge and Disk}.
pp 21--30

\bibitem[\protect\citeauthoryear{{Girardi}, {Bertelli}, {Bressan}, {Chiosi},
  {Groenewegen}, {Marigo}, {Salasnich} \& {Weiss}}{{Girardi}
  et~al.}{2002}]{GIR02}
{Girardi} L.,  {Bertelli} G.,  {Bressan} A.,  {Chiosi} C.~.,  {Groenewegen}
  M.~A.~T.,  {Marigo} P.,  {Salasnich} B.,    {Weiss} A.,  2002, \aap, 391, 195

\bibitem[\protect\citeauthoryear{{H{\" a}fner}, {Evans}, {Dehnen} \&
  {Binney}}{{H{\" a}fner} et~al.}{2000}]{HAF00}
{H{\" a}fner} R.,  {Evans} N.~W.,  {Dehnen} W.,    {Binney} J.,  2000, \mnras,
  314, 433

\bibitem[\protect\citeauthoryear{{Hammersley}, {Garz{\' o}n}, {Mahoney}, {L{\'
  o}pez-Corredoira} \& {Torres}}{{Hammersley} et~al.}{2000}]{HAM00}
{Hammersley} P.~L.,  {Garz{\' o}n} F.,  {Mahoney} T.~J.,  {L{\'
  o}pez-Corredoira} M.,    {Torres} M.~A.~P.,  2000, \mnras, 317, L45

\bibitem[\protect\citeauthoryear{{Hammersley}, {Garzon}, {Mahoney} \&
  {Calbet}}{{Hammersley} et~al.}{1994}]{HAM94}
{Hammersley} P.~L.,  {Garzon} F.,  {Mahoney} T.,    {Calbet} X.,  1994, \mnras,
  269, 753

\bibitem[\protect\citeauthoryear{{He}, {Whittet}, {Kilkenny} \& {Spencer J
  ones}}{{He} et~al.}{1995}]{HE95}
{He} L.,  {Whittet} D.~C.~B.,  {Kilkenny} D.,    {Spencer J ones} J.~H.,  1995,
  \apjs, 101, 335

\bibitem[\protect\citeauthoryear{{Ibata} \& {Gilmore}}{{Ibata} \&
  {Gilmore}}{1995a}]{IBA95a}
{Ibata} R.~A.,  {Gilmore} G.~F.,  1995a, \mnras, 275, 591

\bibitem[\protect\citeauthoryear{{Ibata} \& {Gilmore}}{{Ibata} \&
  {Gilmore}}{1995b}]{IBA95b}
{Ibata} R.~A.,  {Gilmore} G.~F.,  1995b, \mnras, 275, 605

\bibitem[\protect\citeauthoryear{{Kuijken}}{{Kuijken}}{1996}]{KUI96}
{Kuijken} K.,  1996, in IAU Symp. 169: Unsolved Problems of the Milky Way {Is
  There a Bulge Distinct from the Bar?}.
pp 71--+

\bibitem[\protect\citeauthoryear{{Kumar} \& {Riffert}}{{Kumar} \&
  {Riffert}}{1997}]{KUM97}
{Kumar} P.,  {Riffert} H.,  1997, \mnras, 292, 871

\bibitem[\protect\citeauthoryear{{L{\' e}pine} \& {Leroy}}{{L{\' e}pine} \&
  {Leroy}}{2000}]{LEP00}
{L{\' e}pine} J.~R.~D.,  {Leroy} P.,  2000, \mnras, 313, 263

\bibitem[\protect\citeauthoryear{{L{\' o}pez-Corredoira}, {Cabrera-Lavers},
  {Gerhard} \& {Garz{\' o}n}}{{L{\' o}pez-Corredoira} et~al.}{2004}]{LOP04}
{L{\' o}pez-Corredoira} M.,  {Cabrera-Lavers} A.,  {Gerhard} O.~E.,    {Garz{\'
  o}n} F.,  2004, \aap, 421, 953

\bibitem[\protect\citeauthoryear{{L{\' o}pez-Corredoira}, {Hammersley},
  {Garz{\' o}n}, {Cabrera-Lavers}, {Castro-Rodr{\'{\i}}guez}, {Schultheis} \&
  {Mahoney}}{{L{\' o}pez-Corredoira} et~al.}{2001}]{LOP01}
{L{\' o}pez-Corredoira} M.,  {Hammersley} P.~L.,  {Garz{\' o}n} F.,
  {Cabrera-Lavers} A.,  {Castro-Rodr{\'{\i}}guez} N.,  {Schultheis} M.,
  {Mahoney} T.~J.,  2001, \aap, 373, 139

\bibitem[\protect\citeauthoryear{{Lopez-Corredoira}, {Garzon}, {Mahoney} \&
  {Calbet}}{{Lopez-Corredoira} et~al.}{1997}]{LOP97}
{Lopez-Corredoira} M.,  {Garzon} F.,  {Mahoney} T.,    {Calbet} X.,  1997,
  \mnras, 292, L15

\bibitem[\protect\citeauthoryear{{Mackay}, {McMahon}, {Beckett}, {Gray},
  {Ellis}, {Firth}, {Hoenig}, {Lewis}, {Medlen}, {Parry}, {Pritchard} \&
  {Sabbey}}{{Mackay} et~al.}{2000}]{MAC00}
{Mackay} C.~D.,  {McMahon} R.~G.,  {Beckett} M.~G.,  {Gray} M.,  {Ellis} R.~S.,
   {Firth} A.~E.,  {Hoenig} M.,  {Lewis} J.~R.,  {Medlen} S.~R.,  {Parry}
  I.~R.,  {Pritchard} J.~M.,    {Sabbey} C.~S.,  2000, in Proc. SPIE Vol. 4008,
  p. 1317-1324, Optical and IR Telescope Instrumentation and Detectors,
  Masanori Iye; Alan F. Moorwood; Eds. {CIRSI: the Cambridge infrared survey
  instrument for wide-field astronomy}.
pp 1317--1324

\bibitem[\protect\citeauthoryear{{Mathis}}{{Mathis}}{1990}]{MAT90}
{Mathis} J.~S.,  1990, \araa, 28, 37

\bibitem[\protect\citeauthoryear{{Matteucci}, {Romano} \& {Molaro}}{{Matteucci}
  et~al.}{1999}]{MAT99}
{Matteucci} F.,  {Romano} D.,    {Molaro} P.,  1999, \aap, 341, 458

\bibitem[\protect\citeauthoryear{{McNamara}, {Madsen}, {Barnes} \&
  {Ericksen}}{{McNamara} et~al.}{2000}]{MCN00}
{McNamara} D.~H.,  {Madsen} J.~B.,  {Barnes} J.,    {Ericksen} B.~F.,  2000,
  \pasp, 112, 202

\bibitem[\protect\citeauthoryear{{McWilliam} \& {Rich}}{{McWilliam} \&
  {Rich}}{1994}]{MCW94}
{McWilliam} A.,  {Rich} R.~M.,  1994, \apjs, 91, 749

\bibitem[\protect\citeauthoryear{Merrifield}{Merrifield}{2003}]{MER03}
Merrifield M.~R.,  2003, astro-ph/0308302

\bibitem[\protect\citeauthoryear{{Nakada}, {Onaka}, {Yamamura}, {Deguchi},
  {Hashimoto}, {Izumiura} \& {Sekiguchi}}{{Nakada} et~al.}{1991}]{NAK91}
{Nakada} Y.,  {Onaka} T.,  {Yamamura} I.,  {Deguchi} S.,  {Hashimoto} O.,
  {Izumiura} H.,    {Sekiguchi} K.,  1991, \nat, 353, 140

\bibitem[\protect\citeauthoryear{{Omont}, {Ganesh}, {Alard} \& {The Isogal
  Collaboration}}{{Omont} et~al.}{1999}]{OMO99}
{Omont} A.,  {Ganesh} S.,  {Alard} C.,    {The Isogal Collaboration} 1999,
  \aap, 348, 755

\bibitem[\protect\citeauthoryear{{Persson}, {Murphy}, {Krzeminski}, {Roth} \&
  {Rieke}}{{Persson} et~al.}{1998}]{PER98}
{Persson} S.~E.,  {Murphy} D.~C.,  {Krzeminski} W.,  {Roth} M.,    {Rieke}
  M.~J.,  1998, AJ, 116, 2475

\bibitem[\protect\citeauthoryear{{Picaud}, {Cabrera-Lavers} \& {Garz{\'
  o}n}}{{Picaud} et~al.}{2003}]{PIC03}
{Picaud} S.,  {Cabrera-Lavers} A.,    {Garz{\' o}n} F.,  2003, \aap, 408, 141

\bibitem[\protect\citeauthoryear{{Reid}}{{Reid}}{1993}]{REI93}
{Reid} M.~J.,  1993, \araa, 31, 345

\bibitem[\protect\citeauthoryear{{Sabbey}, {McMahon}, {Lewis} \&
  {Irwin}}{{Sabbey} et~al.}{2001}]{SAB01}
{Sabbey} C.~N.,  {McMahon} R.~G.,  {Lewis} J.~R.,    {Irwin} M.~J.,  2001, in
  ASP Conf. Ser. 238: Astronomical Data Analysis Software and Sys tems X
  {Infrared Imaging Data Reduction Software and Techniques}. p 317--+

\bibitem[\protect\citeauthoryear{{Salaris} \& {Girardi}}{{Salaris} \&
  {Girardi}}{2002}]{SAL02}
{Salaris} M.,  {Girardi} L.,  2002, \mnras, 337, 332

\bibitem[\protect\citeauthoryear{{Sevenster}}{{Sevenster}}{1999}]{SEV99}
{Sevenster} M.~N.,  1999, \mnras, 310, 629

\bibitem[\protect\citeauthoryear{{Sevenster} \& {Kalnajs}}{{Sevenster} \&
  {Kalnajs}}{2001}]{SEV01}
{Sevenster} M.~N.,  {Kalnajs} A.~J.,  2001, \aj, 122, 885

\bibitem[\protect\citeauthoryear{{Stanek} \& {Garnavich}}{{Stanek} \&
  {Garnavich}}{1998}]{STA98}
{Stanek} K.~Z.,  {Garnavich} P.~M.,  1998, \apjl, 503, L131+

\bibitem[\protect\citeauthoryear{{Stanek}, {Mateo}, {Udalski}, {Szymanski},
  {Kaluzny} \& {Kubiak}}{{Stanek} et~al.}{1994}]{STA94}
{Stanek} K.~Z.,  {Mateo} M.,  {Udalski} A.,  {Szymanski} M.,  {Kaluzny} J.,
  {Kubiak} M.,  1994, \apjl, 429, L73

\bibitem[\protect\citeauthoryear{{Unavane} \& {Gilmore}}{{Unavane} \&
  {Gilmore}}{1998}]{UNA98}
{Unavane} M.,  {Gilmore} G.,  1998, MNRAS, 295, 637+

\bibitem[\protect\citeauthoryear{{van Loon}, {Gilmore}, {Omont}, {Blommaert},
  {Glass}, {Messineo}, {Schuller}, {Schultheis}, {Yamamura} \& {Zhao}}{{van
  Loon} et~al.}{2003}]{VAN03}
{van Loon} J.~T.,  {Gilmore} G.~F.,  {Omont} A.,  {Blommaert} J.~A.~D.~L.,
  {Glass} I.~S.,  {Messineo} M.,  {Schuller} F.,  {Schultheis} M.,  {Yamamura}
  I.,    {Zhao} H.~S.,  2003, \mnras, 338, 857

\bibitem[\protect\citeauthoryear{{Weinberg}}{{Weinberg}}{1992}]{WEI92}
{Weinberg} M.~D.,  1992, \apj, 384, 81

\bibitem[\protect\citeauthoryear{{Wyse}, {Gilmore} \& {Franx}}{{Wyse}
  et~al.}{1997}]{WYS97} 
{Wyse} R.,  {Gilmore} G.,    {Franx} M.,  1997, ARAA, 35, 637+

\end{thebibliography}

\label{lastpage}

\end{document}